\documentclass[review]{elsarticle}

\usepackage{lineno,hyperref}
\usepackage[bottom]{footmisc}
\usepackage{cleveref}
\usepackage{color}
\usepackage{todonotes}
\usepackage{soul}
\usepackage{xspace}
\usepackage{caption}
\usepackage{subcaption}
\usepackage{booktabs}
\usepackage{multirow}
\usepackage{array}
\usepackage{makecell}
\usepackage[T1]{fontenc}
\usepackage{comment}
\usepackage{rotating}
\usepackage{placeins}
\usepackage[pdf]{graphviz}
\hypersetup{colorlinks = true,linkcolor = blue,anchorcolor =red,citecolor = blue,filecolor = red,urlcolor = red,pdfauthor=author}

\modulolinenumbers[5]

\newcommand{\reva}[2]{#1 #2} 
\newcommand{\revb}[2]{#1 #2}

\journal{Journal of Information and Software Technology}

\begin{document}

\begin{frontmatter}

\title{An initial Theory to Understand and Manage Requirements Engineering Debt in Practice}
% \title{Understanding Requirements Engineering Debt}
    
\author[bth]{Julian Frattini\corref{correspondingauthor}}
\cortext[correspondingauthor]{Corresponding author}

\author[bth]{Davide Fucci}

\author[bth,fortiss]{Daniel Mendez}

\author[richmond,salvador]{Rodrigo Spinola}
\author[novisad]{Vladimir Mandi\'c}
\author[novisad]{Neboj\v sa Tau\v san}
\author[kau]{Muhammad Ovais Ahmad}
\author[bth]{Javier Gonzalez-Huerta}

\address[bth]{Blekinge Institute of Technology, Valhallvägen 1, 37140 Karlskrona, Sweden}
\address[fortiss]{fortiss GmbH, Guerickestraße 25, 80805 Munich, Germany}
\address[richmond]{Virginia Commonwealth University, 23284 Richmond, United States}
\address[salvador]{Salvador University, 41820 Salvador, Brazil}
\address[novisad]{University of Novi Sad, Dr Zorana \DJ in\dj ića 1,
21102 Novi Sad, Serbia}
\address[kau]{Universitet Karlstad, Universitetsgatan 2, 651 88 Karlstad, Sweden}

\begin{abstract}
    \textbf{Context:} Advances in technical debt research demonstrate the benefits of applying the financial debt metaphor to support decision-making in software development activities.
    % Technical Debt research focuses on code, architecture, design, and testing. 
    Although decision-making during requirements engineering has significant consequences, the debt metaphor in requirements engineering is inadequately explored.
    \textbf{Objective:} We aim to conceptualize how the debt metaphor applies to requirements engineering by organizing concepts related to practitioners' understanding and managing of requirements engineering debt (RED).  
    % Mapping the debt metaphor to RE requires an informed definition of \textit{requirements engineering debt} (RED).
    \textbf{Method:} We conducted two in-depth expert interviews to identify key requirements engineering debt concepts and construct a survey instrument. We surveyed 69 practitioners worldwide regarding their perception of the concepts and developed an initial analytical theory.
    \textbf{Results:} We propose a RED theory that aligns key concepts from technical debt research but emphasizes the specific nature of requirements engineering. In particular, the theory consists of 23 falsifiable propositions derived from the literature, the interviews, and survey results. 
    \textbf{Conclusions:} The concepts of requirements engineering debt are perceived to be similar to their technical debt counterpart. Nevertheless, measuring and tracking requirements engineering debt are immature in practice. Our proposed theory serves as the first guide toward further research in this area.
\end{abstract}


\end{frontmatter}

%\linenumbers

\section{Introduction}
\label{sec:intro}

% Context 
Over the past decade, Technical Debt (TD) has emerged as a popular research area~\cite{rios2018tertiary} and a useful model to support decision-making in software development~\cite{siebra2012managing}.
The concept of TD leverages the financial metaphor of debt to model trade-offs between short-term benefits and long-term effects. 
For example, shortcuts in software development reduce time-to-market but require increasingly expensive future refactoring to address quick-and-dirty solutions. 

The TD metaphor extends to a range of software engineering (SE) activities.
The tertiary study by Rios et al.~\cite{rios2018tertiary} shows that the concept of debt is applicable to and researched in the fields of---amongst others---software architecture, design, and quality assurance, effectively establishing concepts such as Architecture Debt, Design Debt, Test Debt, and others.
In Requirements Engineering (RE), Ernst~\cite{ernst2012role} proposed an initial application of the debt metaphor---i.e., \textit{requirements engineering debt} (RED).
Subsequently, Lenarduzzi and Fucci~\cite{lenarduzzi2019towards} provided a vision for a broader definition of how debt can be accumulated in requirements engineering activities.

% Problem
Despite these initial contributions that recognized RED as one of the different types of \revb{debt}{, gaps in research persist}.
First and foremost, we lack a deep understanding of how and to what extent RED is perceived and addressed in the industry.
Similar to its code-related counterpart, RED is a complex concept characterizing practical phenomena that cannot exist in a vacuum and that can only be observed in practical contexts.
Consequently, understanding and further refining the concept of RED requires studying it in the real world. 
In contrast, TD research has progressed towards establishing such an empirically grounded understanding through a strong community of researchers and practitioners exemplified by the InsighTD project\footnote{\url{http://www.td-survey.com/}}.
% , which aims at continuously improving the empirically grounded understanding of TD through a family of global surveys.  

% Why is a problem worth solving
Without a comparable understanding of RED from a practical viewpoint, the relevance of future research is questionable.
In this paper, we argue that RED as decision-making support in practice needs to consider practitioners' perspectives.
The necessity to fill the research gap and advance RED research is exacerbated by the impact RE has on subsequent software development activities~\cite{wagner2019status}.
\revb{It}{is desirable not only to detect and manage these issues as early as possible but also to understand their consequences properly}.
We argue that RED has the potential to model such a relationship.
% Understanding RED can support awareness and decision-making to prevent or reduce these scaling impacts.
Given the viability of the debt metaphor at code level to support decision-making, studying the metaphor in the RE context potentially yields even larger benefits in terms of, for example, cost avoidance.
\revb{Since}{defects introduced during the RE phase of the software development process tend to scale approximately by a factor of 10~\cite{boehm1988understanding}, the cost avoidance through supported decision-making in the form of RED has great potential.}

% Step towards solving
\revb{We}{performed a three-step empirical study as a starting point for addressing the existing gaps.
First, we conducted expert interviews to derive an initial set of themes related to the debt metaphor in the RE context.
Second, we developed a questionnaire based on the interview results and conducted an online survey to gauge practitioners' perceptions of the identified RED concepts. 
Third, we used the 69 survey responses from practitioners to develop a descriptive, evidence-based theory of RED serving as the conceptual foundation for further research.}

% Contribution
We make the following contributions:
\begin{enumerate}
    \item \textbf{Conceptualization}: A refined map of key concepts from the TD metaphor to the RE domain. % JUF: I would remove this from the contributions. I believe we have agreed on the fact that the mapping of TD to RE already exists, we only contribute the empirical perspective. DMZ: Changed to "elaborate" as we take existing discussions further. Maybe not the strongest of our contributions, but maybe still justified to be listed?
    \item \textbf{Analytical theory}: An analytical theory that describes the notion of RED based on our empirical studies. The theory supports decision-making in RE by leveraging the debt metaphor.
    \item \textbf{Data}: The material necessary to replicate the interview and survey studies.\footnote{Accessible in our replication package, currently at \url{https://www.dropbox.com/s/zbz7zjnwn62x87a/ured-replication.zip?dl=0}. The data will be stored permanently on Zenodo upon acceptance.}
\end{enumerate}
To the best of our knowledge, this is the first empirical work explicitly tackling the understanding of RED in practice.

% Paper organization
\revb{In}{the rest of the manuscript, \Cref{sec:related} presents the current state of TD and RED research. 
\Cref{sec:study} presents the study design for the interview study, the online survey, and the subsequent theory development. 
The results of these studies are presented in~\Cref{sec:results}.
Finally, \Cref{sec:discussion} discusses the results, including their limitations, and maps them to the existing TD theory in. 
\Cref{sec:conclusion} concludes with an outline of future work.}

\section{Background and Related Work}
\label{sec:related}

\revb{\Cref{sec:related:terminology}}{discusses the fundamental terminology used in the context of TD and \Cref{sec:related:red} further introduces related work}.

\subsection{Technical Debt Terminology}
\label{sec:related:terminology}

The term technical debt (TD) has been originally coined by Cunningham to describe short-term decisions at the expense of long-term consequences~\cite{cunningham1992wycash}.
In particular, the financial concept of debt---where the total cost is composed of a fixed principal and a time-dependent interest---has been used to describe how these short-term decisions accrue additional cost at an interest rate that ultimately needs to be paid back on top of the principal to remediate the debt~\cite{kruchten2019managing}.
\revb{The}{notion of TD sparked a productive research field, including a dedicated conference venue\footnote{\url{https://conf.researchr.org/series/TechDebt}} and related concepts like the converse \textit{technical credit} introduced by Berenbach et al., which is defined as the ``investment in the engineering, designing and constructing of software or systems over and above the minimum necessary effort, in anticipation of emergent properties paying dividends at a later date.''~\cite{berenbach2014technical}}

A secondary study by Li et al.~\cite{li2015systematic} synthesized existing TD literature to identify the most commonly agreed-upon concepts of the metaphor.
TD is composed of items, which are ``a unit of TD in a software system''~\cite{li2015systematic}. 
TD can have several causes and effects, where the latter subdivide into consequences, symptoms, and value:
\begin{itemize}
    \item Consequence: ``the influences of incurring TD on the software system''~\cite{li2015systematic}.
    \item Symptom: indicators for the incurred TD.
    \item Value: ``the potential benefit of incurring TD''~\cite{li2015systematic}.
\end{itemize}
Further concepts like bankruptcy, which ``happens when the part of the software system which contains TD is no longer viable to support the development and a complete rewrite, and a new platform are needed''~\cite{o2010technical}, are less commonly explored but still relevant to TD research.

\subsection{TD research in RE}
\label{sec:related:red}

Among the fifteen types of TD identified by Rios et al.~\cite{rios2018tertiary}, two types are related to RE: documentation debt and requirements debt. 

Documentation debt represents the problems in documentation artifacts produced during various development stages, including requirements specifications.
This limits the scope of documentation debt to requirements \textit{artifacts}, which are only one aspect of RE. 

Rios et al.~\cite{rios2020hearing} conducted a study focusing on documentation debt.
Surveying 39 practitioners and interviewing experts revealed generalized causes, effects, prevention, and repayment practices specific to documentation debt.
While the prevention and repayment practices are more focused on code-related documentation (e.g., ``[c]omment the code'' as a prevention and ``[k]eep the documentation updated'' as a payment practice~\cite{rios2020hearing}), the identified causes of debt are also relevant to RED (e.g., ``[d]eadline'' and ``[i]naccurate time estimate''~\cite{rios2018most}). 

Barbosa et al.~\cite{barbosaorganizing} specifically investigate a subset of documentation debt in the domain of requirements engineering, which they frame \textit{requirements and requirements documentation debt} (R2DD).
Using an existing data set containing survey responses of 78 practitioners to the global InsighTD survey~\cite{rios2020practitioners}, the authors extracted causes, effects, prevention, and repayment practices relevant to requirements documentation.
The resulting occurrences of the observed concepts are presented in~\Cref{tab:conceptsliterature}.

\begin{table}[ht!]
\footnotesize
    \centering
    
    \begin{tabular}{l l l}
	\toprule
		\makecell{\textbf{Concepts}}		& \makecell{\textbf{Documentation debt \cite{rios2020practitioners}}}	& \makecell{\textbf{Requirements debt \cite{barbosaorganizing}}} \\ \hline
		
		Causes								& \makecell{\textit{Deadline},\\ The company does not give\\ value to documentation,\\ Non-adoption of\\ good practices,\\ Inaccurate time estimations,\\\textit{Inappropriate planning}.\\ }      & \makecell{\textit{Deadline},\\ Not effective project\\ management,\\ Change in requirement,\\ \\ \textit{Inappropriate planning,}\\ High turnover of the team. }  \\ \hline
		
		Effects								& \makecell{ \textit{Low maintainability},\\ \textit{Delivery delay},\\ \textit{Rework},\\ \textit{Low external quality},\\ Inadequate, non-existing or\\outdated  documentation.}      & \makecell{ \textit{Delivery delay},\\ \textit{Rework},\\ Financial loss,\\ \textit{Low external quality},\\ \textit{Low maintainability}.}  \\ \hline
		
		\makecell{Prevention\\ practices}	&  \makecell{Comment the code,\\ Create tutorials on how to\\ fill the documentation,\\ Define process and good\\ practices for documentation,\\ Define roles concerning the\\ documentation process,\\ Document the project since\\ it begins.}     &  \makecell{Well-defined requirement,\\ Follow the project\\ planning, \\ Follow the well-defined\\ project process, \\ Well-defined scope\\ statement,\\ Good allocation of\\ resources in the team.} \\ \hline
		
		\makecell{Repayment\\ practices}	& \makecell{Adopt TD payment\\ prioritization criteria,\\ Keep the documentation\\ updated, \\ Review outdated\\ documentation.}  & \makecell{Code refactoring,\\ \\Monitor and control\\ project activities,\\ Design refactoring,\\ Investing effort on \\TD repayment activities, \\ Changing project scope.}  \\ \hline
		
	\end{tabular}
    \caption{The occurrences of TD concepts identified in InsighTD surveys, ranked based on their occurrences (most occurring first). Italic text represents concepts identified in both studies.}
    \label{tab:conceptsliterature}
\end{table}

These two previous studies~\cite{rios2020hearing,barbosaorganizing} constitute the empirical body of knowledge of TD applied to RE. 
\revb{Even}{though their scope is} limited to documentation, their results are relevant to RED and we discuss their alignment in \Cref{sec:discussion:mapping}.

Ernst~\cite{ernst2012role} presents one of the first definitions of technical debt in requirements as ``the distance between the optimal solution to a requirements problem and the actual solution, with  respect  to  some  decision space.''
The distance results from decisions trading immediate gains for future costs, which corresponds to the notion of intentional TD.
Similarly to unintentional TD, the distance can also increase due to unforeseen or unintended changes in the context of the requirements problem~\cite{ernst2012role}.
Accordingly, the interest on the debt is the rate of increase of such distance~\cite{ernst2012role}.

This definition exceeds the perspective of requirements documentation and considers general \textit{requirements problems}.
\revb{Whereas}{debt applying to the requirements documentation only refers to requirements artifacts, the scope of RED extends to the whole RE process.}
However, the definition of RED is still limited to only one specific requirements problem---i.e., what Lenarduzzi and Fucci~\cite{lenarduzzi2019towards} refer to as \textit{mismatch implementation}.
The latter extends the definition of RED by including two additional requirements problems related to \textit{stakeholder discovery}---i.e., the debt accrued by involving only part of the stakeholders in the RE process, and \textit{requirements artifact} smells---i.e., the debt accrued due to quality violation in the requirements specification, such as using ambiguous language. 

More recently, a systematic literature review by Melo et al.~\cite{MFL22} covered the causes of RED as well as methods to identify and manage it.
They identified 66 primary studies that, although not explicitly focusing on RED, contained instances of investigations addressing activities related to RE, for instance, in design and documentation. 
Their results show 33 causes for RED divided according to their level of intentionality and 16 strategies to identify and manage RED, including nine issues arising in such a process.
Moreover, inspired by TD research, they provide metrics to assess RED principal, interests, and payback.

Through the application of the TD metaphor first to requirements documentation~\cite{barbosaorganizing,rios2020hearing} and later to requirements engineering in general~\cite{ernst2012role,lenarduzzi2019towards,MFL22}, the research community established a conceptual foundation of RED. However, an empirical perspective on RED in practice is still a gap that we address in this study.

\section{Study Design}
\label{sec:study}

In this study, we answer the following research questions:
\begin{enumerate}
    \item RQ1: How do practitioners understand RED?
    \item RQ2: How do practitioners manage RED?
\end{enumerate}
Similar to previous studies related to TD in RE~\cite{barbosaorganizing,rios2020hearing}, we are interested in understanding the concepts of RED and how to manage it.
In contrast to previous studies on RED~\cite{ernst2012role,lenarduzzi2019towards,MFL22}, we focus on the view of practitioners.

Our study is divided into three stages: (1) expert interviews, (2) a questionnaire-based online survey, and (3) the inference of a theory.
The interview study is a first and in-depth investigation of practitioners' perspectives on RED, while the online survey scales up the involvement of practitioners. 
As a final step, we synthesize our results into an initial theory of RED.
This theory serves as a foundation for supporting decision-making in RE utilizing the debt metaphor and provides a starting point for empirically grounded follow-up research.

We present the overall study design in~\Cref{sec:study:pipeline} and explain the three individual steps in more detail in \Cref{sec:study:interview,sec:study:survey,sec:study:theory}. 

\subsection{Research Pipeline}
\label{sec:study:pipeline}

\begin{figure}[ht]
    \centering
    \includegraphics[width=\textwidth]{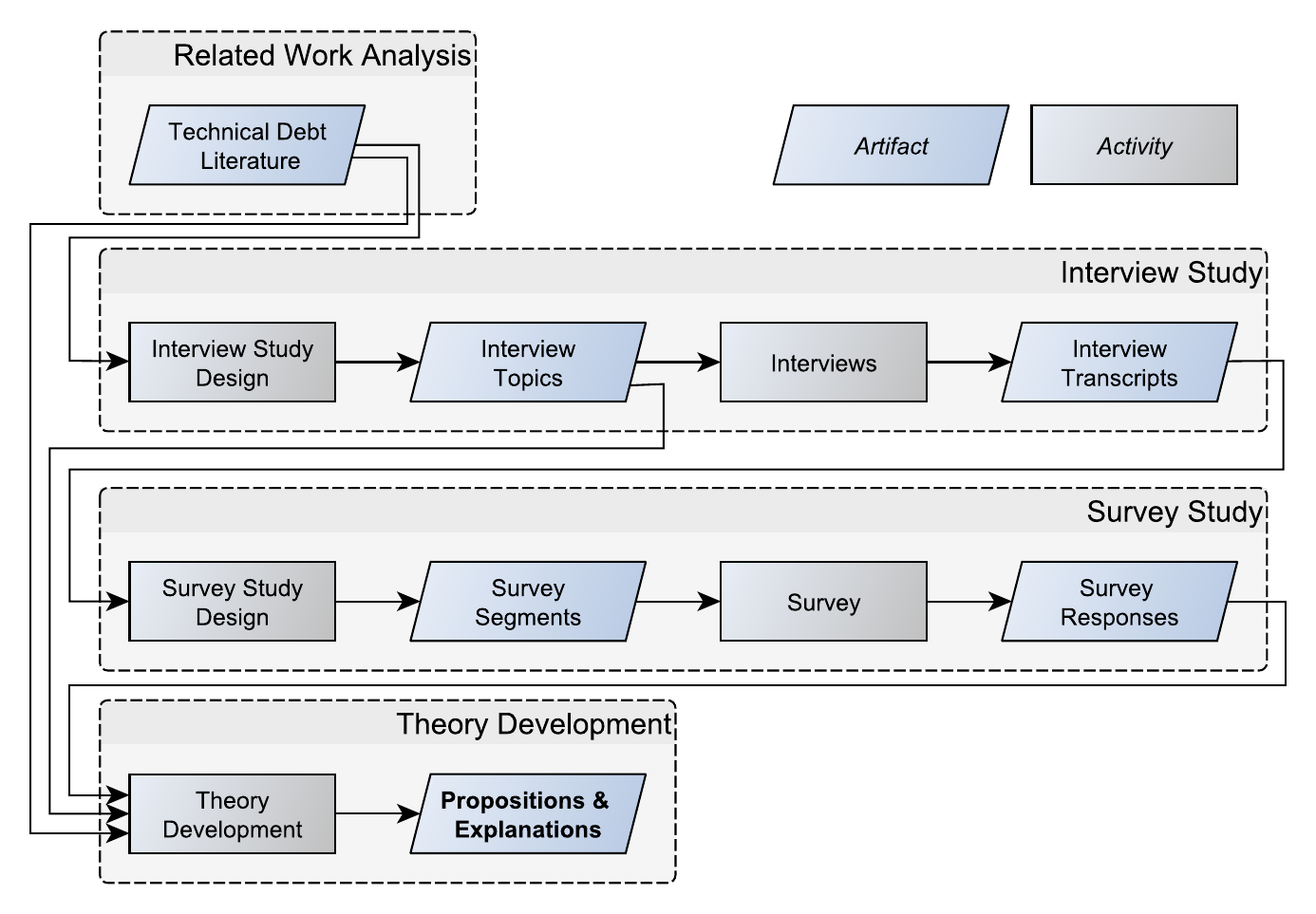}
    \caption{Research process pipeline.}
    \label{fig:study:design}
\end{figure}

We visualize the overall research process in~\Cref{fig:study:design}.
We initially identified common concepts from the TD literature and mapped them to the RE domain.
This mapping was aligned with the theoretical contributions to RED literature (e.g.,~\cite{ernst2012role,lenarduzzi2019towards}).
The concepts guided the design of the interview study and resulted in a set of themes to structure the interview protocol.
We used the transcripts produced during the interviews as input to design the online survey instrument.
We used the responses to the survey and the interview topics to derive the propositions and explanations in our theory.
Where applicable, we used existing literature to support our explanations.

Table~\ref{tab:conceptmap} lists all relevant concepts used within the TD domain and connects them to their respective counterparts in the three steps of this study.

\begin{table}[ht]
    \centering
    \caption{Mapping of TD concepts from literature to interview themes and survey segments.}
    \begin{tabular}{l|l|l}
    \toprule
        \makecell{\textbf{TD Literature}}                        &  \textbf{Interview topics}                           & \textbf{Survey segment}          \\                       \midrule
        Debt Item~\cite{li2015systematic}                                  & Debt Item                                 & ---                                           \\
        Value~\cite{curtis2012estimating}                                   & Value-cost tradeoff                       & Value                                          \\
        \makecell{Principal \& interest~\cite{curtis2012estimating}}      & Cost structure                            & ---                                             \\
        Bankruptcy~\cite{o2010technical}                                  & Cost Structure                            & Bankruptcy                                    \\
        Impact~\cite{brown2010managing}                                      & TD Properties                             & Propagation                                   \\
        Symptoms~\cite{li2015systematic}                                 & TD Properties                           & Symptoms                                       \\
        Causes~\cite{rios2018most}                                        & Causes                                 & Causes                                         \\
        Effects~\cite{rios2018most}                                       & Effects                                  & Effects                                       \\
        Awareness~\cite{ernst2015measure}                             & Awareness                             & Intentionality                                 \\
        Management~\cite{kruchten2019managing}                          & Management                               & Remediation                                    \\
        Tracking~\cite{guo2011tracking,seaman2011measuring}              & Monitoring            &  \makecell{Detecting, Measuring, Tracking}    \\ \bottomrule
    \end{tabular}
    \label{tab:conceptmap}
\end{table}

\subsection{Interview Study}
\label{sec:study:interview}
The goal of the interviews was to collect guiding themes and constructs for creating the survey questionnaire (see~\Cref{sec:study:survey}). 
We used the key informant technique~\cite{Mar96} to collect valuable evidence from experts on a specific topic. 
In particular, we selected experts with the following characteristics, the participants (1) have a role in the \reva{RE}{} community which exposes them to the kind of information we sought, (2) are knowledgeable in the field of RE and TD, and (3) are willing to communicate their knowledge. 

\paragraph{Participants} 
The first participant is the co-founder of a company that provides services, including training, in the area of RE quality improvement with seven years of practical experience.
The other two are product managers in a large software service provider with over 20 years of experience.
All participants are also active researchers in the areas of RE and SE. 
\reva{We}{used convenience sampling~\cite{baltes2022sampling} as we recruited the interview participants from our personal networks.
After noticing convergence of the guiding themes and constructs elicited by all three interview participants, we deemed the data collection of the interview study sufficiently complete.}

\paragraph{Interview protocol}
% interview design: changes to the concept map (@Davide)
We selected the topics for the interview study---and, accordingly, the interview script---around themes mapped and adapted from the TD literature (see~\Cref{tab:conceptmap}).
In particular, we (1) extended the concept of \textit{Value} to include an explicit discussion of its trade-off with costs; (2) organized the concepts of \textit{Principal\&Interest} and \textit{Bankruptcy} around the theme \textit{Cost structure}; (3) addressed \textit{Impact} and \textit{Symptoms} together in the \textit{TD properties} theme;
% iv) used the \textit{Scope} theme to discuss not only the \textit{Impact} on artefacts---as defined by Li et al.~\cite{li2015systematic}---but also on other development phases;
(4) used the term \textit{Monitoring}---as suggested by Rios et al.~\cite{rios2018tertiary}---to include the several activities (e.g., detecting, measuring) that fall under the \textit{Tracking} concept.

% interview process
Before the interviews, we communicated our research objectives to the participants.
We conducted the interviews remotely over video between 2020-04-21 and 2020-06-25.
The second author acted as the interviewer, and the first and the third author supported the process by taking notes and prompting visualization material to aid the main interviewer.

\begin{sidewaysfigure}
    \centering
    \includegraphics[width=\textwidth]{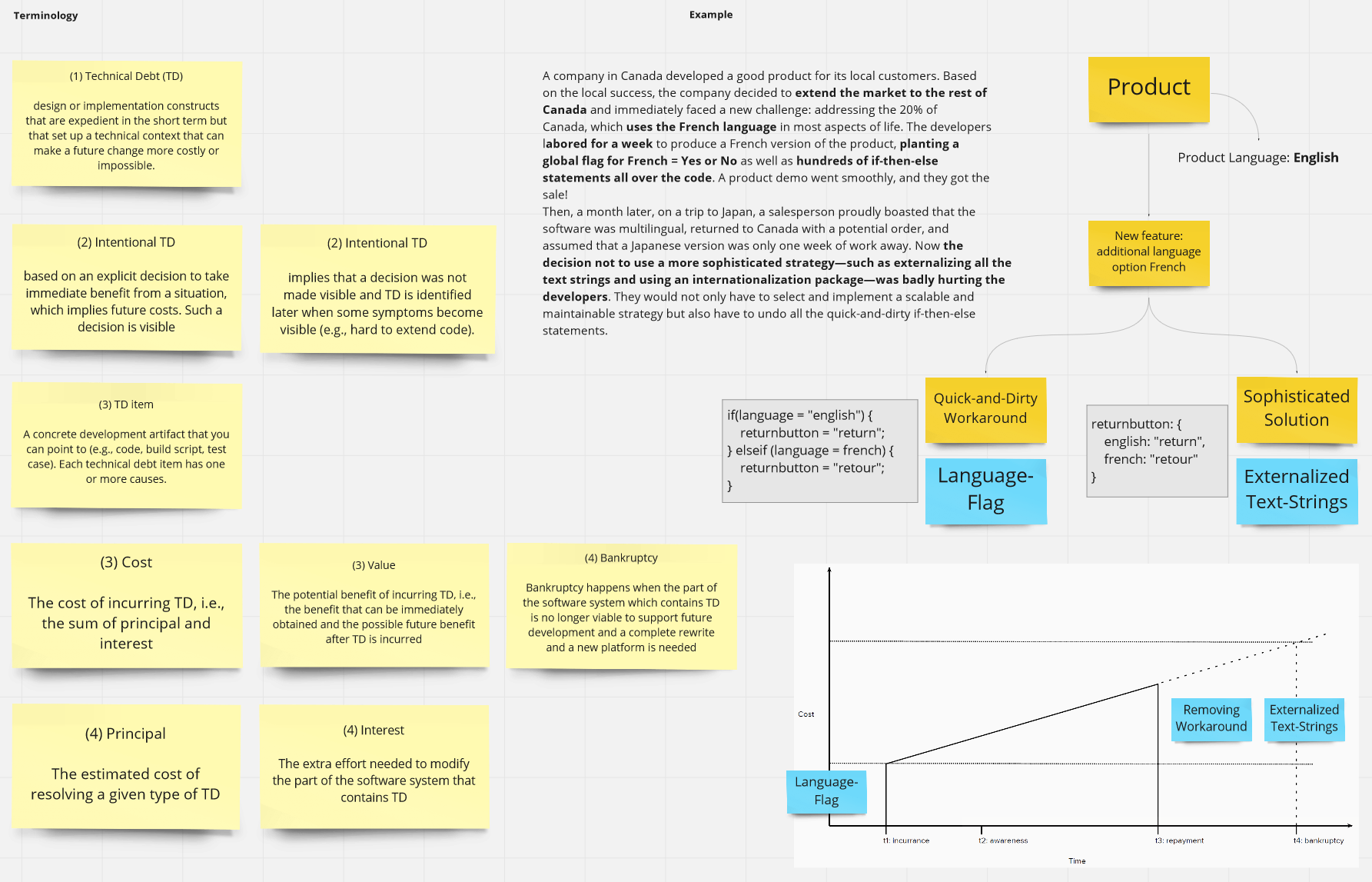}
    \caption{\revb{Interview}{aid visualizing concepts and examples.}}
    \label{fig:interview:aid}
\end{sidewaysfigure}

We used a semi-structured interview approach.
In particular, we focused on understanding whether and how the participants apply the TD metaphor to RE in practice (for instance, how they understand a debt item in RE or what is the cost of a debt item in RE) and what strategies they use to manage it.
We used a shared online whiteboard as support to drive the interviews, clarify the constructs, and provide examples to elicit participants' perspectives as shown in~\Cref{fig:interview:aid}.
Each interview lasted approximately 70 minutes and was voice recorded.
We then used the transcription of the recording for thematic analysis~\cite{CD11}.

\begin{figure}
    \centering
     \includegraphics[width=.85\textwidth]{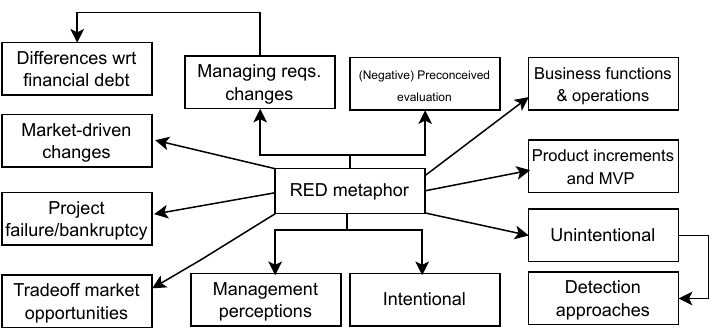}

    \caption{Additional themes from the interviews after open and axial coding.}
    \label{fig:themes}
\end{figure}

\paragraph{Analysis}
The first two authors independently read the entire transcript of each interview and conducted, during a first analysis instance, deductive coding~\cite{saldana2021coding} against interview topics to identify, for example, causes, effects, and properties of RED. 
In a second analysis instance, the same authors performed inductive coding to avoid missing relevant aspects related to the RED metaphor \revb{and}{ensure completeness of the coding}.
\reva{In total,}{the two authors generated 114 codes. After comparing individual codes, agreeing on common labels, and performing axial coding, they yield the 13 themes reported in~\Cref{fig:themes}.
For example, under \textit{Project failure/bankruptcy}, we coded the following statement by P2:
``So you have written down stuff and you realize that maintaining this documentation, as it may be, is too expensive. Let's just drop it. That happens.'' as \textit{maintenance}. 
Also, P3 stated: ``You accept too many requirements and at the end you have a very complicated system that tries to enforce a process that nobody wants to use in practice and the market fights against it as they don't want to use it. I think this system is in a bankruptcy situation because of bad requirements engineering.'' which we coded as \textit{gold plating} under the same theme.
The complete codebook is available in the replication package; verbatim quotes were removed due to non-disclosure agreement.}
%After open coding~\cite{saldana2021coding} and agreeing on common label definitions, they performed axial coding.
% The process resulted in 13 themes (see~\Cref{fig:themes}) which the third author checked for consistency.

\subsection{Online Survey}
\label{sec:study:survey}

We developed an online questionnaire based on the themes identified when analyzing the interviews to scale up the insights gained through the practitioners' perspectives on RED.
% Design: target population and study objects
\paragraph{Target Population and Sampling} The target population of our survey consists of practitioners generally engaged with requirements---whether as part of an explicit requirements engineering phase or not.
Here, we applied convenience sampling. 
We reached out to our target population by contacting experts in RE who are well-connected to the industry in different parts of the world.
We did so via our personal contacts, networks, and associations. 
We complemented this outreach with posts on social media (such as Twitter). 
Participants acted as multipliers, and each multiplier was encouraged to distribute the survey invitation in their respective networks.

% Implementation and execution: using unipark, timeframe
\paragraph{Questionnaire} We designed the survey segments and the questions contained in each segment based on the evaluation of the interview transcripts (see~\Cref{tab:conceptmap}).
In addition, we (1) embedded the concept of \textit{debt items} into adjacent concepts (causes and consequences) to make them more tangible; (2) disregarded \textit{cost structure} as the interview revealed that the current lack of quantification for RED makes an assessment of principal and interest unlikely; instead, we focused on \textit{bankruptcy} as a concept underrepresented in literature; (3) emphasized \textit{propagation} based on the interview discussions of \textit{impact}; (4) specified \textit{awareness} into \textit{intentionality} (i.e., either intentional or unintentional), which we also consider to be more suitable for analysis in a survey; (5) concretized management and tracking into detecting, measuring, tracking, and remediation. 
The survey protocol is available in \ref{sec:apx:survey}.

The questionnaire consists of 33 closed and 10 open-ended questions, of which six closed and one open-ended questions were demographic questions.
We used closed-ended questions to capture the practitioners' perceptions regarding RED concepts, which we later included in the theory---for example, the proposition about the most dominant type of causes for RED (P3 in \Cref{tbl:proposition:1}).
Two open-ended questions let respondents provide a rationale for their answers to closed-ended questions.
In the other seven open-ended questions, respondents can consider more complex answers requiring them to provide examples based on their experiences.
The questionnaire was implemented using a web-based tool and distributed via mail to our contacts who acted as multipliers and posted on social media.
We collected a total of 69 complete survey responses in the timeframe between 10.05.2021 and 03.02.2022.

\paragraph{Analysis Procedure} We analyzed the quantitative data from the survey using descriptive statistics and visualizations. 
% We consider the sample size to be suitable to answer our research questions.
However, the sample distribution does not enable meaningful statistical analysis (e.g., a comparison between countries and domains).
We focused on qualitative data analysis to answer our research questions. 
In particular, we use open coding~\cite{saldana2021coding} to derive explanations for existing concepts as well as concepts not previously considered.

\subsection{Theory Development}
\label{sec:study:theory}
Finally, we use the responses from the survey results to build an initial theory of RED.
Considering the terminological and conceptual adoption of theory building and evaluation for Software Engineering~\cite{wieringa2014design, sjoberg2008building}, we constructed an analytical theory focusing on a model that captures a set of propositions that can later serve as a basis to define (falsifiable) hypotheses. 

While building the theory, we draw from a similar experience in the context of the \textit{Naming the Pain in Requirements Engineering} initiative\footnote{\url{www.napire.org}}~\cite{wagner2019status,fernandez2017naming}.
However, rather than starting by populating our theory first with constructs and propositions from literature and then corroborating it with the survey results (as done in~\cite{wagner2019status}), we use the survey results to build our theory \revb{bottom-up}{} and then add explanations, where suitable, by drawing from existing evidence from the literature.
We selected this approach due to the scarcity of RED-specific literature. 

\section{Results}
\label{sec:results}

\revb{\Cref{sec:results:interview}}{presents the results from the interview study, \Cref{sec:results:survey} the results of the online survey, and \Cref{sec:results:theory} the results from the theory development.}

\subsection{Interview Study}
\label{sec:results:interview}

The themes derived from coding the interview transcripts showed how existing TD concepts are understood in RE and revealed new concepts that are underrepresented in existing TD literature but are relevant to RED.
While the elements of the debt metaphor (e.g., debt item, causes, effects) mostly aligned with existing research, the concepts \textit{intentionality} and \textit{propagation} emerged as particularly relevant to RED.
Discussions on \textit{managing RED} revealed some relevant deviations of RED from TD. 
\reva{The}{interview study confirmed the applicability of the debt metaphor to the RE domain and added valuable, additional concepts to consider in the subsequent survey study.}

\paragraph{Elements of the debt metaphor} The interview participants perceived elements like \textit{debt item}, \textit{principal \& interest}, \textit{value}, \textit{symptoms} and \textit{effects} similarly to how they are understood in TD.
One interview participant reports ``taking a shortcut in the RE process'' as the value of RED, which matches its equivalent definition in TD.
Further, the participant reports that ``the impact of mistakes in requirements engineering is like a hundred, a thousand times higher than fixing it right away,'' which aligns with an exponential \textit{interest} of the debt item cost structure.
The \textit{effects} of RED are described by one interview participant using a three-tier hierarchy~\cite{hilty2005information}---``if you have a problem in your specification, everybody who is using your specification can make a mistake [due to the first order impact]. Then, second-order impact [affects] the people who use those artifacts that are created based on your requirements. Test cases are based on your requirements, system code is based on your requirements. Everybody who is using those can be impacted by that.''
The third tier encompasses everything affected by the deployed product, which maximizes the cost of remediating the debt items.
This description aligns with the cost structure of TD and emphasizes the cascading consequences that RED can have.
The \textit{causes} of RED are described by one interview participant as time-related ``project pressure'' and ``lack of understanding of what's good engineering.''
According to these statements, RED must not be limited to documentation but has to consider organizational as well as personal aspects.

\paragraph{Managing RED} The interview participants emphasized that managing RED is not widely established.
One interview participant reported ``I've just seen one company where I looked into these things and they have been systematically documenting the results of the requirements reviews. And most other companies, it's more like, <<let's fix the issues.>> That's all we want to do.'' 
Tool support for managing RED specifically is lacking, with one participant suggesting that ``[w]hat you need is an automatic analysis that not just checks your requirements at a specific point in time, but over a period of time.''
Another interview participant reports that ``the toolset JIRA, Kanban boards and that---they are okay. However, it is more about practice and culture rather than tools.'' 
As ``you need to keep track of your debt,'' the need for both awareness and an explicit approach to managing RED becomes apparent.

\paragraph{Intentionality} Whether debt is accrued intentionally or unintentionally is an important attribute of TD as well as RED.
The interview participants reported that RED is most commonly accrued unintentionally due to a lack of diligence, \revb{i.e.,}{being oblivious to a better alternative}.
In requirements engineering, ``one of the main unintentional debts is probably that you don't ask enough stakeholders'', as reported by one of the interview participants.
However, RED can also be accrued intentionally---i.e., a debt item is intentionally created with explicit awareness.
This usually happens as a consequence of negligence, as one interview participant reports---``I see it sometimes when we, as consultants, especially when we do audits, we come in and tell people, <<Hey, you need to fix these ten fundamental structural problems>> and nobody fixes them.'' 
Another interview participant confirms that intentional debt is often accrued due to active prioritization---``[the debt item] gets intentionally de-prioritized because other things are just more important.''
One interview participant reports that ``[the companies] often do not track the intentional [debt]. The unintentional [debt] is something you have to detect and then track, and you have to invest some effort and time to make sure that you detect it.''

\paragraph{Propagation} A less-explored concept in TD and RED literature is the \textit{propagation} of debt from one debt item to other artifacts, causing them to also accrue debt~\cite{zabardast2022assets}. 
Interview respondents reported that one risk associated with RED is that its impact on artifacts in the scope of RE (e.g., specifications or use cases) can cause debt for other artifacts outside the scope of RE.
One interview participant reported the potential scope of this affecting even the final product, when saying ``this can be [...] up to very high levels that you say, <<Actually we can throw the whole product away>>'', which corresponds to the bankruptcy of the product.
The propagation of RED can be mitigated by domain knowledge. 
``[I]f you have a great team who knows the domain, who knows the application, and knows everything about the system, who are users of your system themselves, it's all there. They're experts actually for the application. Then you have an okay chance that they will still do a good job. In that case, fixing that technical debt just is within the scope of the requirements.'', one interview participant reported.

\subsection{Survey Results}
\label{sec:results:survey}

In this section, we present the results of the survey's quantitative and qualitative questions. While analyzing the qualitative questions, we identified four main themes and several associated sub-themes, summarized in \Cref{tbl:codes}.
The sub-themes are not mutually exclusive---for example, the \textit{time pressure} sub-theme can apply to more than one of the main themes. 

\subsubsection{Demographics}

\begin{figure}[h]
    \centering
    \includegraphics[width=0.8\textwidth, trim=0 0 0 19, clip]{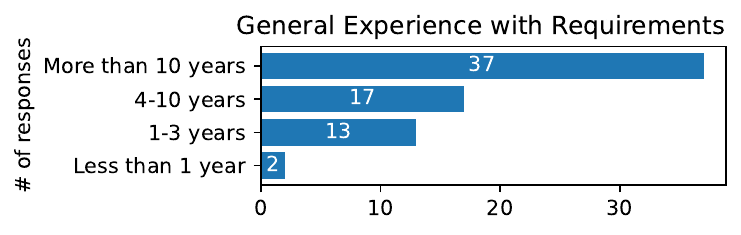}
    \caption{Experience of participants}
    \label{fig:survey:demo:experience}
\end{figure}

\begin{figure}[h]
    \centering
    \includegraphics[width=\textwidth, trim=0 0 0 19, clip]{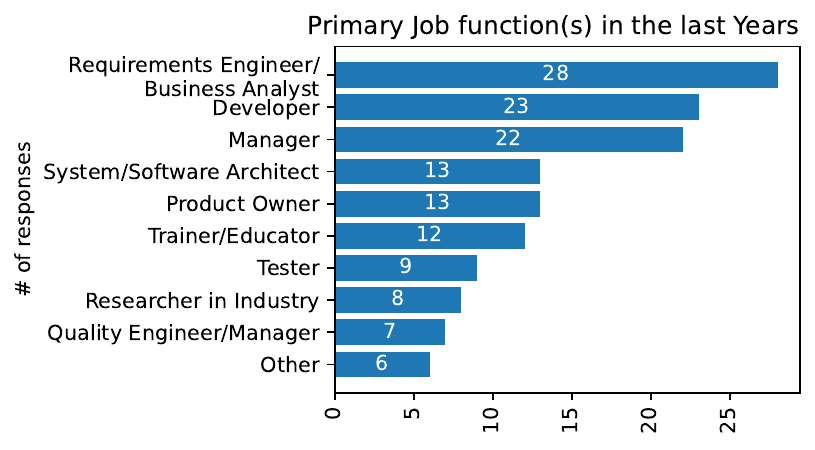}
    \caption{Occupation of Respondents}
    \label{fig:survey:demo:job}
\end{figure}

\begin{figure}[h]
    \centering
    \includegraphics[width=0.9\textwidth, trim=0 0 0 19, clip]{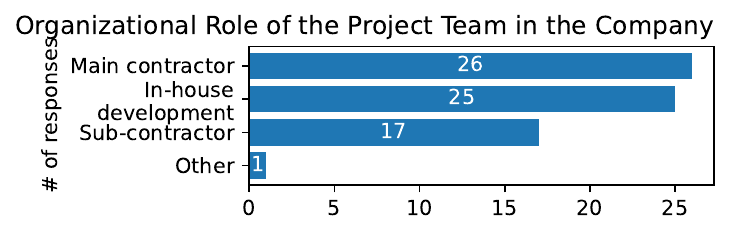}
    \caption{Role within the organization}
    \label{fig:survey:demo:role}
\end{figure}

\begin{figure}[h]
    \centering
    \includegraphics[width=\textwidth, trim=0 0 0 19, clip]{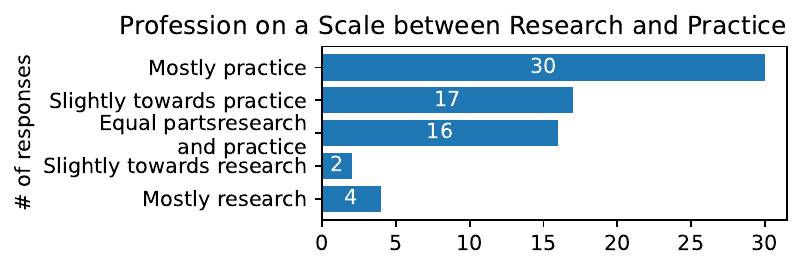}
    \caption{Scale of Profession}
    \label{fig:survey:demo:profession}
\end{figure}

\begin{figure}[h]
    \centering
    \includegraphics[width=\textwidth, trim=0 0 0 19, clip]{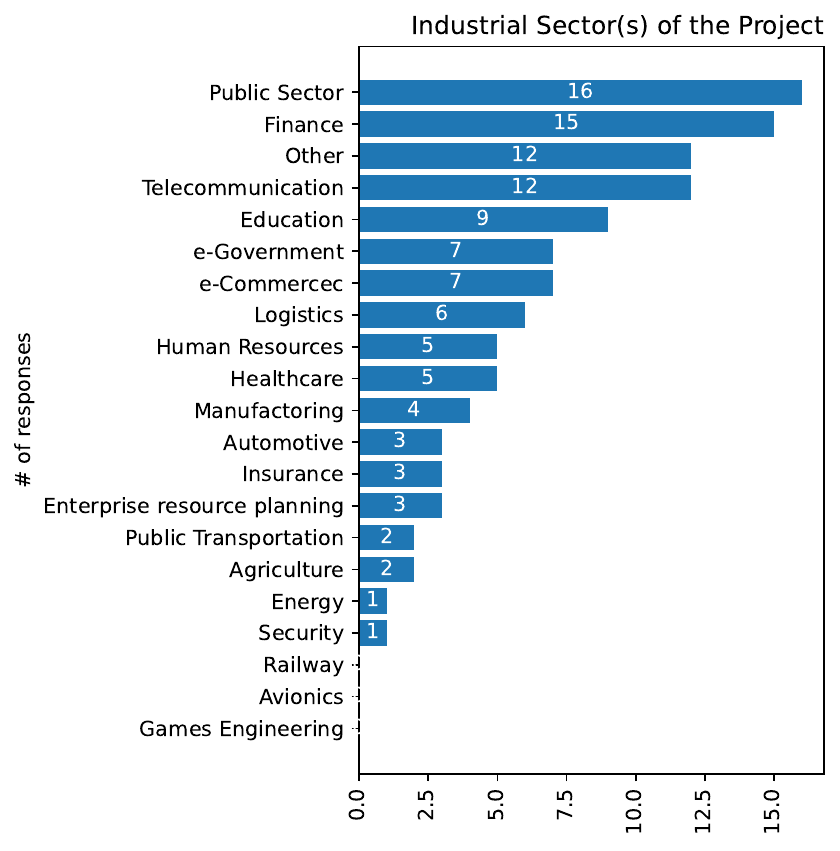}
    \caption{Main sector of the respondent's company}
    \label{fig:survey:demo:sector}
\end{figure}

\begin{figure}[h]
    \centering
    \includegraphics[width=0.8\textwidth, trim=0 0 0 19, clip]{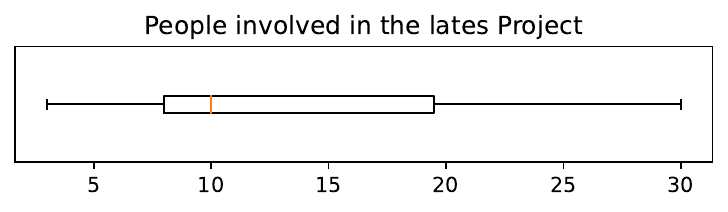}
    \caption{Project team sizes}
    \label{fig:survey:demo:size}
\end{figure}

The survey was accessed by 125 potential participants and completed by 69 of them, which yields a completion rate of $55.2\%$. 
For the results, we considered only responses to the entire questionnaire.
The majority of participants (n=37) have more than 10 years of experience in requirements engineering, as shown in \Cref{fig:survey:demo:experience}. 
The primary job functions (see \Cref{fig:survey:demo:job}) of the respondents are in operational roles, such as requirements engineers (n=28) and developers (n=23), as well as managerial (n=22).
The majority of the respondents are performing their job functions (see \Cref{fig:survey:demo:role}) as the main contractor of a company  (n=26) or as part of in-house development (n=25).
The majority of respondents rank themselves as rather practice- than research-oriented, as shown in \Cref{fig:survey:demo:profession}.
The represented industrial sectors include a variety of domains as shown in \Cref{fig:survey:demo:sector} with the exception of the \textit{railway}, \textit{avionics}, and \textit{games engineering} domains. 
The respondents are involved in projects with a median team size of 10 individuals, although some involve up to 30 persons (see \Cref{fig:survey:demo:size}).
Out of the 69 participants, 40 are based in South and 5 in North America, 23 in Europe, and one in Asia. 

\begin{sidewaystable}
\centering
\footnotesize
 \caption{Summary of the (sub-)themes emerged from the the free-text answers to the survey.}
\begin{tabular}{lllr}
\toprule
\multicolumn{1}{c}{\textbf{Theme}}                   & \textbf{Sub-theme}                & \textbf{Description} & \textbf{Occurrences} \\ \midrule
\multirow{5}{*}{Factors influencing RED}       & Time pressure        &      A deadline for release a product or service, or their new version   &     28        \\
                                               & Competition          & Information gathered from external stakeholders in the same market         &       13       \\
                                               & Reputation               & Internal customers' complaints and requests             &   5           \\
                                               & Economic aspect                &      Budget adjustments, for example due to loss of sales       &   19           \\ \midrule
\multirow{5}{*}{Assessing and remediating RED} & Time saving        &      Assessment of debt is based on the amount of time/effort saved       &        25     \\
                                               & Customer involvement &   The amount of customers, and which customers are impacted             &       23       \\
                                               &  Agile development&         The process models prioritizes other activities than debt assessment and remediation     &         11     \\
                                               & Review sessions      &       Need for ad-hoc review sessions to assess and remediated debt      &          7    \\
                                               & Traceability         &           Links between artifacts support RED assessment   &       3       \\ \midrule
\multirow{2}{*}{Process and artefacts in RED}  & Agile development                &       Artefacts typical of Agile (e.g., user stories)   &   8           \\
                                               & Rework               & The process includes extra rework due to RED            &      8        \\ \midrule
\multirow{2}{*}{Intentionality}                & Time pressure        &        A deadline for \reva{releasing}{} a product or service, or their new version      &        11      \\
                                               & Skill/knowledge      &     Lack of skills, knowledge, or due diligence      &        9      \\  \bottomrule
\end{tabular}
 \label{tbl:codes}
\end{sidewaystable}

\subsubsection{Causes} 
\begin{figure}
    \centering
    \begin{subfigure}{.3\textwidth}
        \centering
        \includegraphics[width=\textwidth]{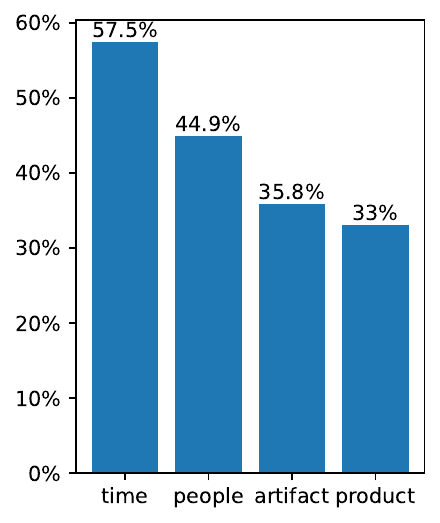}
        \caption{Causes}
        \label{fig:survey:causes}
    \end{subfigure}
    \begin{subfigure}{.3\textwidth}
        \centering
        \includegraphics[width=\textwidth]{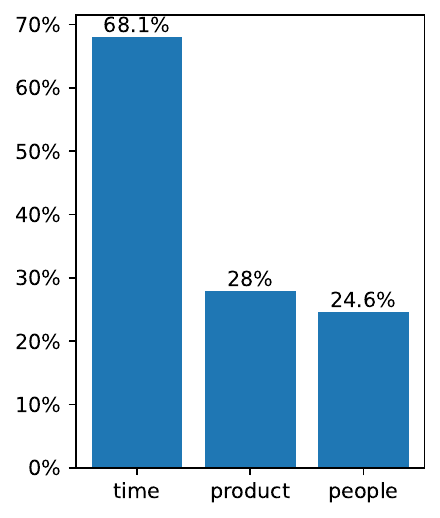}
        \caption{Values}
        \label{fig:survey:values}
    \end{subfigure}
    \begin{subfigure}{.3\textwidth}
        \centering
        \includegraphics[width=\textwidth]{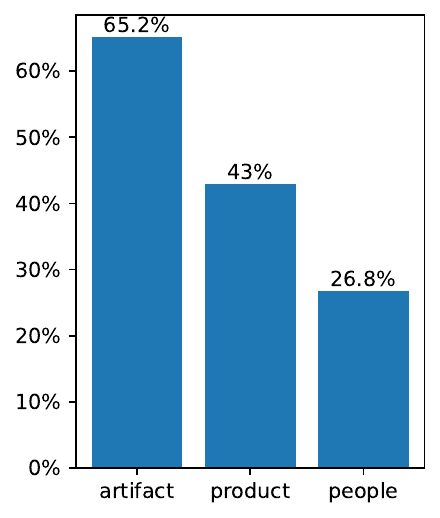}
        \caption{Symptoms}
        \label{fig:survey:symptoms}
    \end{subfigure}
    \caption{RED \textit{Causes}, \textit{Values}, and \textit{Symptoms} indicated by respondents. The y-axis reports average agreement with possible answers in a specific category reported on the x-axis.}
    \label{fig:survey:concepts}
\end{figure}

The first two authors manually categorized the RED causes that emerged from the interview study into \textit{requirements artifact-}, \textit{time-}, \textit{people-}, and \textit{product-}related.
The full list of potential causes and their categorization can be found in the Appendix in Table~\ref{tab:apx:groups:cause}.
In the survey, we ask respondents to indicate one or more causes for RED from this list.
\Cref{fig:survey:causes} shows the support for the categories of \textit{Causes} averaged over all answers within a single category. 
For example, the causes in the time-related category, on average, were agreed upon by 57.5\% of the survey participants.
Respondent showed a strong agreement with time-related answers like \textit{time pressure to deliver a feature} and \textit{time pressure to finalize the (systematic) requirements specification}.
People-related answers received the next-strongest support, with \textit{lack of domain knowledge} and \textit{lack of communication between stakeholders} being the most prominent.
Requirements artifact- and product-related answers share comparable support as valid causes of RED.
\textit{Requirements not documented}---\revb{a property}{of the requirements artifacts potentially causing costly rework due to the lack of a tangible document to refer to}---is the most agreed upon artifact-related cause \revb{for}{RED}, corroborating the findings of Wagner et al.~\cite{wagner2019status} that requirements incompleteness remains one of the most critical aspects of requirements engineering.
Answers in the product-related category include \textit{technical complexity of the project}, which is still deemed influential but less so compared to the aforementioned causes.
The respondents also point specifically to the lack of skills resulting in accruing RED items. 
For example, one survey respondent commented on their answer that ``decisions are made by product and business people without the background to realize their consequences. They know they are cutting corners and creating the debt on purpose (to speed up delivery usually).''
Others pointed out the lack of knowledge regarding what the professional figure of a requirements engineer entails, as people without such knowledge unintentionally introduce debt.
For example, one respondent commented: ``Organizations have been trying to document less than necessary and having people engage in conversations and meetings to cover for the gaps, creating RE debt intentionally.  When organizations are asked to document their requirements properly, they often don't know how to do it or create low-quality requirements, creating debt unintentionally.''

\begin{figure}
    \centering
    \includegraphics[width=\textwidth]{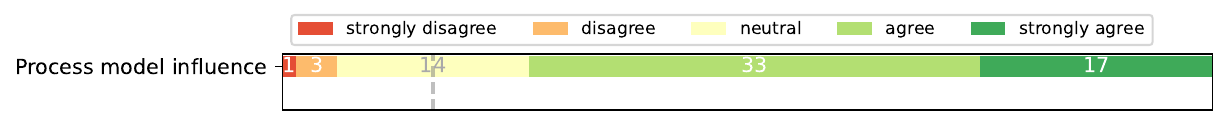}
    \caption{Influence of process model on the introduction of RED.}
    \label{fig:survey:processmodel}
\end{figure}

The last comment stresses the role of the utilized process model and its influence in causing the accumulation of RED.
When specifically asked about this, the respondents tend to agree that the process model impacts the likelihood of introducing RED (see \Cref{fig:survey:processmodel}). 
They acknowledge that characteristics of Agile (e.g., focus on the customer, lightweight requirement format, communication) impose trade-offs when managing RED.
For example, one respondent mentioned, ``I believe that in the Agile model we can achieve faster results, but in compensation for the lack of formalization of requirements, it can end up generating requirements engineering debt.'' and also ``Agile projects suffer more requirements engineering debt due to the prioritization of good team communication instead of detailed requirements documentation.''

\begin{figure}
    \centering
    \includegraphics[width=\textwidth]{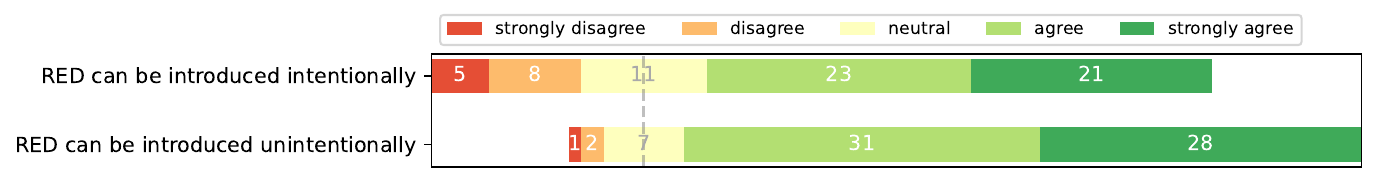}
    \caption{Agreement regarding the intentionality of introducing RED.}
    \label{fig:survey:intentional}
\end{figure}

Furthermore, participants agreed that requirements engineering debt could be introduced both intentionally and unintentionally, as shown in \Cref{fig:survey:intentional}.
We identified themes related to \textit{time pressure} and \textit{skills and knowledge} as the main drivers for the decision to introduce RED. 
Some respondents mention time pressure as the reason to intentionally introduce RED at the very initial phase of RE, such as requirements gathering (e.g., ``Requirements engineering debt may be intentionally introduced if interested parties reduce the time required to gather requirements.''); whereas others acknowledge that time pressure is the results of external factors, such as customers demands (e.g., ``it is needed to accelerate the development of a feature (or fix) to please a customer. In these cases, RED can be introduced intentionally. The attention point is: It needs to be paid later.'').

The analysis of the answers to qualitative questions revealed that companies consider several more factors when accruing RED. 
Some of these are the benefits against which debt is traded off. 
% The dominant themes that emerged deal with \textit{time pressure}---e.g., one survey respondent reported \textit{``Focusing on a faster way to market the product makes us skip many steps in the (RE) process.''}. % this seems redundant now that it is embedded into the merged results
For example, themes such as \textit{beating the competition} emerged.
As a survey respondent reported, ``If we will [lose] a big opportunity of product launching before another company, then there is a value in introducing RED.'', and---connected to it---\textit{reputation}, especially when fulfilling legal obligations (e.g., ``we simply have to focus on requirements that are considered as a potentially high risk of fines, in order to keep the company reputation'').
Another important theme is the \textit{economic aspect} which needs to be saved from parts of the RE process and allocated to activities that can push the product on the market faster.

\subsubsection{Value} 
% A set of potential values for RED became apparent when analyzing the results of the interview (see~\Cref{sec:study:interview}).
The first two authors manually categorized a list of values that emerged from the analysis of the interview (see \Cref{tab:apx:groups:value}).
\Cref{fig:survey:values} shows the average support for each category.

Time-related value like \textit{faster time to market} receives the largest amount of agreement among all potential values for accruing requirements engineering debt.
However, only 17\% of the survey respondents report explicitly calculating the value of requirements engineering debt (Q12).  

Most respondents did not pinpoint specific artifacts impacted by RED but agreed that RED cost is perceived in terms of the \textit{rework} such artifacts undergo. 
For example, one respondent commented: ``I believe it [RED] generates a great deal of rework. I've participated in projects where we had to redo several features due to lack of knowledge of the requirements on the part of the responsible stakeholder.''
Other respondents mentioned that RED impacts the rework of other artifacts down the line, such as source code, \revb{which}{aligns with established findings by Boehm et al.~\cite{boehm1988understanding}}. 
For example, they said: ``[paying back RED has] Additional costs on re-design, refactoring of the code.''

\subsubsection{Symptoms} 
\begin{figure}
    \centering
    \includegraphics[width=\textwidth]{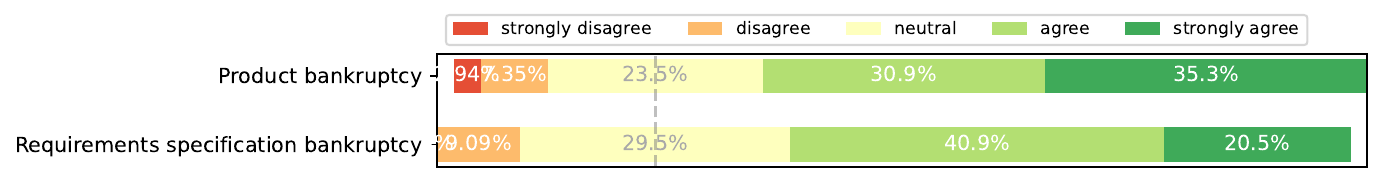}
    \caption{Agreement with the statement ``Requirements engineering debt can cause bankruptcy.''}
    \label{fig:survey:bankruptcy}
\end{figure}

\begin{figure}
    \centering
    \includegraphics[width=\textwidth]{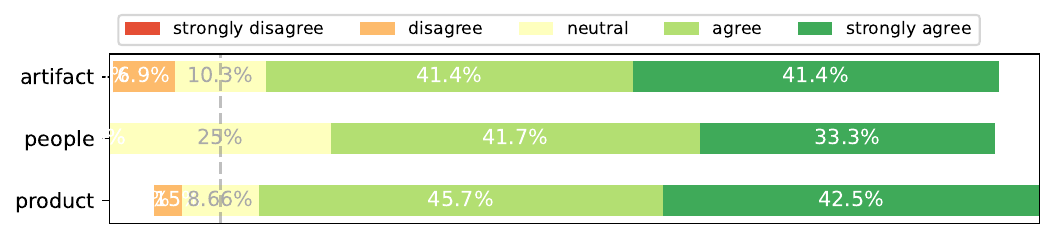}
    \caption{Agreement with the statement, ``The following symptoms of RED can lead to bankruptcy.'' grouped by categories}
    \label{fig:survey:symptoms-bankruptcy}
\end{figure}

The first two authors manually categorized the symptoms which we obtained from the analysis of the interview study  (see~\Cref{tab:apx:groups:symptoms}). 
\Cref{fig:survey:symptoms} shows the support the categories received on average.
Requirements artifact-related symptoms received the strongest support overall, as survey participants agree upon \textit{incompleteness of requirements} being the strongest symptom of requirements engineering debt.
Product-related symptoms like \textit{additional operational cost of the product} and \textit{not implementing requirements} follow, while people-related symptoms like \textit{not involving all relevant stakeholders} receive the least support.
In addition, respondents agreed with the statement that requirements engineering debt could cause bankruptcy for products and requirements specifications, respectively (see~\Cref{fig:survey:bankruptcy}). 
However, the support for the bankruptcy of the product is greater.
\Cref{fig:survey:symptoms-bankruptcy} shows which ones, among the items the respondents reported as symptoms of RED, are likely to indicate impending bankruptcy.

\subsubsection{Propagation} 
\begin{figure}
    \centering
    \includegraphics[width=\textwidth]{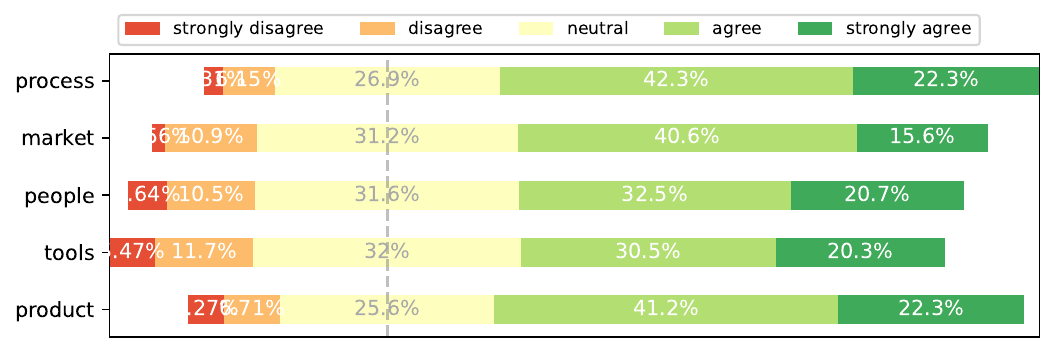}
    \caption{Average agreement with the group of context factors impacting the propagation of RED.}
    \label{fig:survey:8-contextfactors}
\end{figure}

\Cref{fig:survey:8-contextfactors} shows the distribution of agreement about which context factor categories---derived from the list by Petersen et al.~\cite{petersen2009context}---increases the likelihood of RED further propagating. 
The figure shows that process- and product-related context factors are perceived to have the strongest influence on the propagation of RED, though the difference between groups is marginal. 
The strong support for the context factors \textit{maturity}, \textit{quality}, and \textit{size} of the product is offset by the comparably weaker support for \textit{type} and \textit{customization} of the product.\footnote{The complete list of context factors and their categories is available in \ref{sec:apx:categories}, \Cref{tab:apx:groups:context}}
Therefore, a sub-group of product-related context factors (containing maturity, quality, and size of the product) is dominant over all others.

\subsubsection{Management} We survey four types of interaction with RED items; detecting, measuring, tracking, and remediating RED.
While about one-third of the respondents report explicitly detecting (34\%) and remediating (34\%) RED items, a smaller fraction claims measuring (13\%) and tracking (22\%) them. 

\begin{figure}[ht]
    \centering
    \includegraphics[width=\textwidth]{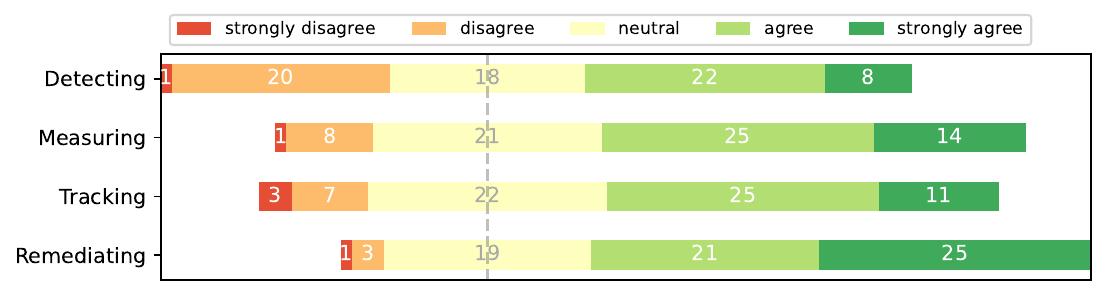}
    \caption{Agreement with the statement, ``\{Action\} is expensive.''}
    \label{fig:survey:9-action-expensive}
\end{figure}

\begin{figure}[ht]
    \centering
    \includegraphics[width=\textwidth]{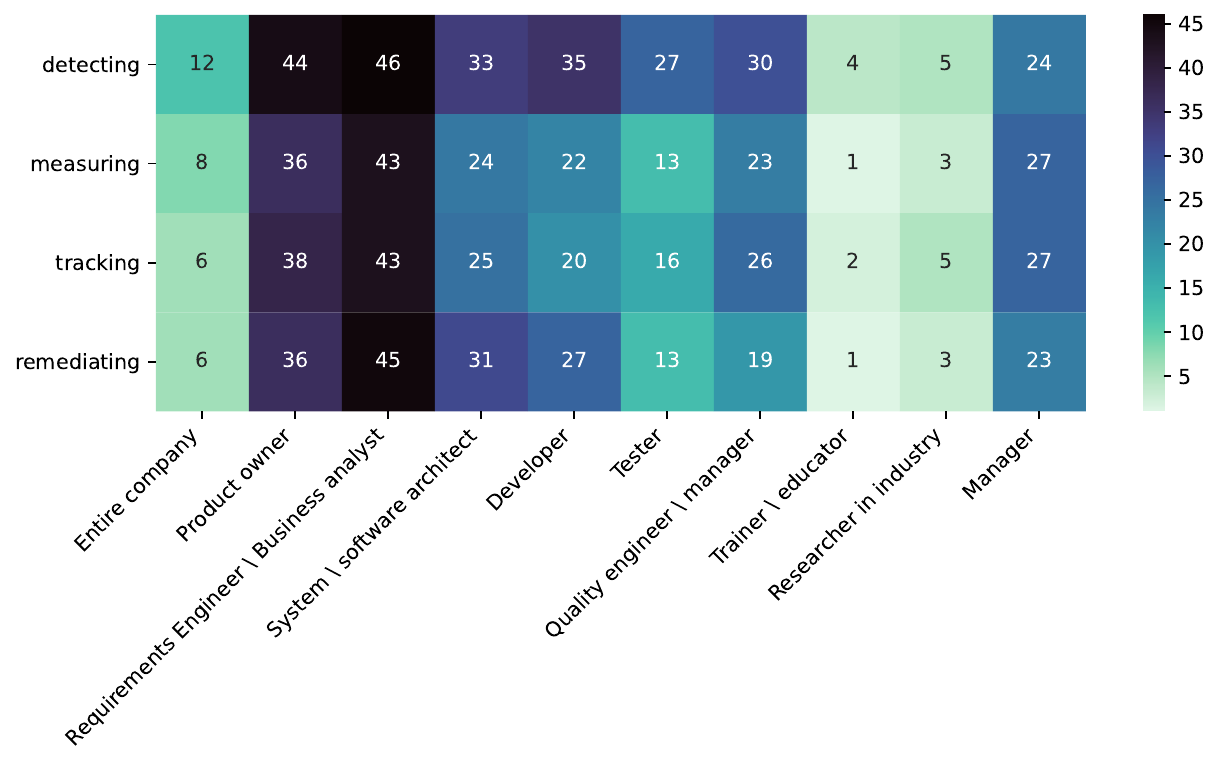}
    \caption{Support for stakeholders being responsible for actions related to managing RED.}
    \label{fig:survey:10-action-resposible}
\end{figure}

Detecting RED is perceived to be comparably the least expensive of the four activities while remediating is the most expensive (see~\Cref{fig:survey:9-action-expensive}).
In general, requirements engineers and product owners are perceived to hold the highest responsibility for interacting with RED, followed by stakeholders in downstream development activities, such as architects, developers, testers, and quality engineers. 
\Cref{fig:survey:10-action-resposible} visualizes the agreement of the survey respondents regarding which role (on the x-axis) is responsible for an activity (on the y-axis). 
For example, 46 respondents agree that a \textit{requirements engineer/business analyst} is responsible for \textit{detecting} RED, while only 12 respondents agree that it should be the responsibility of the \textit{entire company}.

\begin{figure}[ht]
    \centering
    \includegraphics[width=\textwidth]{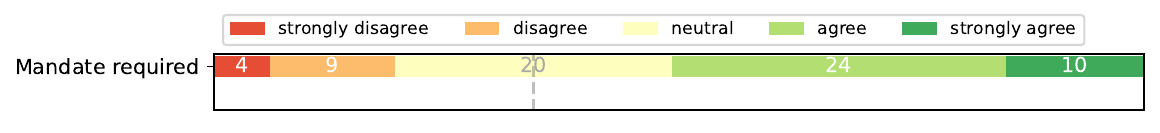}
    \caption{Support for the statement ``Remediating RED requires a mandate.''}
    \label{fig:survey:11-mandate}
\end{figure}

Moreover, 34 out of 69 respondents agree that remediating RED requires a mandate, as shown in~\Cref{fig:survey:11-mandate}.

For most companies that measure RED, its actual value is related to \textit{time}, such as the number of days saved by taking debt in order to release a feature more quickly. 
\revb{These}{measurements apply exclusively to intentional as well as unintentional-but-already-detected RED, as they presuppose an awareness of the debt items.}

Another measure is reported to be the number of \textit{users} (or customers) involved when taking debt, as a survey respondent reported: ``the rule is usually no users, then zero value. So, any time spent on a requirement for an unused feature is debt.''
Others recognize that calculating RED is often difficult, as not only the number of customers but also their expectations need to be taken into account---one respondent reported that ``calculating RED value is often related to customer expectations, so it's hard to do.''
Conversely, \textit{customer dissatisfaction} is perceived as the main ``external'' (i.e., not related to processes or stakeholders within the company) symptom of RED.

Customers and their involvement are also at the center of detecting the presence of unintentional RED.
According to the respondents, some \textit{process models}, such as Agile, are better at supporting RED detection (e.g., one respondent pointed out ``Communication is probably the single most important in my opinion. Agile helps with that since you have frequent sprint cycles and small deliveries with valuable feedback that can help raise awareness of what is going wrong; notice that Agile mainly facilitates communication here.''). 

Internally to a company, the main way to detect RED is to have systematic \textit{review sessions} of the requirement artifacts (e.g., ``Review of product requirement documents by peers/other stakeholders.'', as one respondent put it).
However, the respondents revealed that RED is often detected only during development. 
\revb{This}{is in line with established literature which consequently formulated the need for ``early error detection''~\cite{boehm1988understanding}.}

In this case, the symptom of debt is perceived as the lack of (or hard-to-manage) \textit{traceability} between requirements and other artifacts.
RED items are tracked the same way that technical debt items are tracked, using issue trackers---for example, one survey respondent pointed out ``[RED items] can be tracked as research/investigation issues/tickets similarly to how one might track technical debt items.''
Remediation of the tracked RED items seems to be an ad-hoc process, which relies on specific indications provided by \textit{customers} as indicated by this respondent---``we don't explicitly remediate requirements debt, but we do spend a lot of time redesigning and reworking requirements from users complaints. The key component I see is having actionable feedback on what needs to change as early as possible.''

\subsection{Resulting Theory}
\label{sec:results:theory}

Our RED theory is organized around propositions and explanations.
Propositions and explanations about causes, intentionality, and roles are reported in \Cref{tbl:proposition:1}.
\Cref{tbl:proposition:2} covers symptoms and factors influencing RED, while \Cref{tbl:proposition:3} reports aspects related to managing RED, including detecting, measuring, tracking, and remediating it.
% The support of key literature for the propositions is detailed in \Cref{sec:discussion:mapping}.

\begin{table}[ht!]
\caption{Propositions: Causes, value, intentionality, and roles related to RED}
\footnotesize
\label{tbl:proposition:1}
\begin{tabular}{l p{8.5cm} p{1.6cm}}
\toprule
\textbf{ID} & \textbf{Proposition}                                                              & \textbf{Source} \\ \midrule
P1          & RED items can have artifact-, time-, people-, and/or product-related causes.      & Interviews, Literature \\ \midrule
P2          & RED items can have time-, product-, and/or people-related value                   & Interviews, Literature \\ \midrule
P3          & Time-related causes are most relevant.                                           & Survey \\\midrule
P4          & Time-related value is the most relevant.                                         & Survey \\\midrule
P5          & Agile facilitates the introduction of RED items and at the same time their detection. & Survey \\\midrule
% P5          & The agile software process model can have a positive or negative influence on the introduction of RED. & Survey \\\midrule
P6          & The value of RED is rarely calculated in practice.                                & Survey \\\midrule
P7          & The responsibility of estimating the value of RED lies with requirements engineers or high-level stakeholders. & Survey \\\midrule
P8          & RED can be accrued both intentionally and unintentionally.                            & Survey \\\midrule
P9          & Time-related causes are predominantly intentional, people-related causes are mostly unintentional. & Survey \\\toprule
\textbf{ID} & \textbf{Explanation}                                                              & \textbf{Prop.} \\\midrule
E1          & Among the time-related reasons for accruing RED, beating competitors to market is what incentivizes to accrue RED. & P3, P4 \\\midrule
E1.1        & Companies accrue RED when reputation is at stake, especially when fulfilling legal obligation.  & P3, P4 \\\midrule
E2         & Although Agile lacks formalization, which increases the likelihood of introducing (unintentional) RED, its focus on stakeholders communication and reviews helps uncovering RED. & P5 \\\midrule
% E2          & Although agile lacks formalization, its focus on customer communication helps uncovering RED by involving as many stakeholders as possible. & P5 \\\midrule
E3          & There are causes that are more likely to be intentional and some that are rather unintentional. & P8 \\\midrule
E4          & Time-related causes accrue RED intentionally to achieve the time-saving values. People-related causes accrue RED unintentionally, since it is difficult/hard to pinpoint people-related shortcomings (like lack of diligence) & P9 \\ \bottomrule
\end{tabular}
\end{table}

\begin{table}[ht!]
\caption{Propositions: Symptoms and factors influencing RED.}
\label{tbl:proposition:2}
\scriptsize
\begin{tabular}{l p{8.5cm} p{1.6cm}}
\toprule
\textbf{ID} & \textbf{Proposition}                                                                  & \textbf{Source} \\ \midrule
P10         & RED items can have artifact-, time-, people-, and/or product-related symptoms.        & Interviews, Literature \\\midrule
P11         & Incompleteness of requirements is the strongest symptom of RED.                       & Survey, Literature \\\midrule
P12         & Customer dissatisfaction is a strong symptom of RED.   & Survey \\\midrule
P13         & RED can cause bankruptcy of both the SRS and the product.                             & Survey \\\midrule
P14         & Artifact- and product-related symptoms are most likely to lead to bankruptcy.         & Survey \\\midrule
P15         & Context factors influence the likelihood of propagation of RED.                       & Survey \\\toprule
\textbf{ID} & \textbf{Explanation}                                                                  & \textbf{Propositions} \\ \midrule
E5          & Missing requirements remains one of the strongest indicators for accrued debt in RE~\cite{wagner2019status} & P11 \\ \midrule
E6          & Among the people-related symptoms, customer-related symptoms have two dimensions: not only do they need to be reached out to or involved, but also their relationship maintained. & P12 \\\midrule
E7          & There is a point of cumulative RED that makes a SRS or product unusable, forcing the company to discard the current assets and start anew. & P13 \\\midrule
E8          & Especially properties of the product (size, maturity and quality) under development influence how RED propagates from one item to another. & P15 \\\bottomrule
\end{tabular}
\end{table}

\begin{table}[ht!]
\caption{Propositions: Measuring, assessing, monitoring, and remediating RED.}
\label{tbl:proposition:3}
\scriptsize
\begin{tabular}{l p{8.5cm} p{1.6cm}}
\toprule
\textbf{ID} & \textbf{Proposition}                                                                                  & \textbf{Source} \\ \midrule
P16         & The interactions with RED items are detecting, measuring, tracking, and remediating.                  & Interviews, Literature \\\midrule
P17         & It is more expensive to measure, track, and remediate RED items than detect them.            & Survey \\\midrule
%P18         & RED items are either detected ad-hoc or systematically.                                               & Survey \\\midrule
P18         & RED is more often detected and remediated than tracked or measured                                    & Survey \\\midrule
P19         & RED items are tracked using issue trackers similarly to TD items & Survey \\\midrule
P20         & The remediation strategy is heavily dependent on the customer's perspective on the RED item.          & Survey \\\midrule
P21         & RED is often calculated considering the number of affected customers.                        & Survey \\\midrule
P22         & The responsibility of acting upon RED lies with requirements engineers or high-level stakeholders.    & Survey \\\midrule
P23          & Remediating sometimes requires a mandate.                                                            & Survey \\ \toprule
\textbf{ID} & \textbf{Explanation}                      & \textbf{Propositions}  \\\midrule
E9          & Consequences of RED are often detected during downstream activities involving non-RE stakeholders. & P17 \\\midrule
E10         & RED items can be detected during requirements validation systematically, or ad-hoc by downstream stakeholders & P17 \\\midrule
E11         & While RED is often detected ad-hoc and remediated via systematic review sessions, there is a lack of techniques and tools to track and measure RED. & P18 \\\midrule
E12         & There is a lack of tooling tailored towards RED, forcing companies to adapt other tracking tools for this purpose. There might also not be a need for differentiation. & P19 \\\midrule
E13         & There is a lack of systematic remediation approaches, motivating companies' remediation strategies to depend on customers perspectives. & P20 \\\midrule
E14         & There is no straight-forward, established way of calculating RED as it depends on features/characteristics of customers such as their expectations, which are hard to assess. & P21 \\\midrule
E15         & Requirements engineers who are most involved with the requirements and high-level stakeholders are expected to have a sufficient overview over the accumulated RED. & P22 \\\midrule
E16         & The responsibility of interacting with RED is diffused. The responsibility of detecting RED lies with high-level stakeholders, low-level stakeholders---which would enact the remediation---are hesitant to remediate RED items without a mandate. & P23 \\\bottomrule
\end{tabular}
\end{table}

\subsubsection{Propositions}
We follow a conservative approach without any assumptions on the theoretical robustness in parts of the theory---i.e., the lack of sufficient quantitative data does not allow us to make claims about which propositions are more relevant than others. 
Further, we do not describe context conditions for the individual propositions. 
While such information can be valuable for inferring conditional hypotheses, our data does not allow for a balanced contextualization (e.g., regarding roles or industry sectors) across all propositions.

The theory is descriptive~\cite{sjoberg2008building} as it \textit{describes} how practitioners understand and manage RED.
Future work as outlined in \Cref{sec:conclusion}, especially empirical work focusing on specific propositions, will be necessary to enable a more \textit{prescriptive} theory.

\subsubsection{Explanation}

This empirically grounded theory is aimed to serve as decision support in understanding and managing RED.
\Cref{tbl:proposition:1,tbl:proposition:2} guide the \textit{understanding of RED} by characterizing the concepts contained in the RED metaphor.
P3 and P4 emphasize that RED is mostly connected to time-related aspects, both in the form of cause and value.
The potential value of accruing RED---reducing time to market---is commonly accepted as a sufficient reason for accruing it.
Value is rarely calculated in practice (P6) since it is difficult to measure and track RED in detail, \revb{which}{aligns to prominent research on the challenge of measuring software-related quality~\cite{boehm1976quantitative,boehm1988understanding}}.
Although governance around RED is not clear, organizations avail stakeholders in management positions to estimate the value of RED (P7).

RED can be accrued both intentionally and unintentionally (P8), both bearing individual risks as cautioned during the interviews (see \Cref{sec:results:interview}). 
While unintentional RED can go undetected and have a scaling impact on subsequent software development activities, intentional RED can simply be accepted and actively neglected. 
Different causes are more likely to be intentional; for instance, time-related causes are predominantly intentional, while people-related causes are mostly unintentional. 
The analysis of open-ended survey questions suggests that organizations are hesitant to blame individuals for accruing RED.

Missing or incomplete requirements are the strongest symptom of RED (P11), which aligns with findings in literature~\cite{wagner2019status}, as they escalate to customer dissatisfaction when left untreated (P12). 
This can ultimately result in the bankruptcy of a requirements specification and, in the long run, the product as a whole (P13), which emphasizes the scope of RED impact.

Context factors influence the likelihood of propagation of RED (P15). 
Among commonly known context factors, the size, maturity, and quality of the product under development strongly influence the propagation of debt. 

\Cref{tbl:proposition:3} shows \textit{managing RED}, which is decomposed into detecting, measuring, tracking, and remediating (P16) according to the interview results.
Detecting RED items is less expensive than measuring, tracking, and remediating (P17) since detecting may happen ad-hoc.
\revb{This}{corroborates the generally agreed upon notion of scaling cost models, e.g., the spiral model of software cost~\cite{boehm1988understanding}.}
Among the four activities, measuring and tracking RED are the least common (P18), which is explained by a lack of specific tooling (P19) and the difficulty of properly quantifying RED (P21). 

\section{Discussion}
\label{sec:discussion}

\revb{\Cref{sec:discussion:rqs}}{answers the research questions and discuss the relevance of this research.
In particular, \Cref{sec:discussion:mapping} discusses the results in relation to existing evidence and the general notion of TD. 
Finally, \Cref{sec:discussion:threats} addresses threats to validity.}

\subsection{Answers to the Research Questions}
\label{sec:discussion:rqs}
% The following two subsubsections have a very similar purpose to the "Explanation"-section above. I feel like they have to be differentiated stronger.
\subsubsection{RQ1: How do practitioners understand RED?}
The concepts of RED---e.g., the notion of TD applied to RE---are perceived to be similar to TD.
These concepts consist of causes, items, effects, symptoms, and value. 
Most concepts can be categorized to be artifact-, time-, people-, or product-related (P1, P2). 
Time-related causes and value are most relevant (P3, P4), highlighting the motivation behind accruing RED to be time-savings. 
RED can be accrued intentionally and unintentionally (P7), though certain categories of concepts are strongly associated with either.
For example, time-related causes are predominantly intentional, while people-related causes are mostly unintentional (P9).

A strong symptom of RED encountered \textit{during} the development process is the incompleteness of requirements (P11), whereas customer dissatisfaction (P12) is a post-release symptom. 
RED can ultimately cause the bankruptcy of both the requirements specification and the product (P13), representing a point where the respective entity must be discarded completely.
Context factors, especially the size, maturity, and quality of the product under development, influence the likelihood of RED propagation to other artifacts or processes (P15).

% I am leaving this comment here in the hope that I will stumble upon it again in a further review of this manuscript. 
% Agile is facilitating the introduction of RED item (because lack of formalization), and its removal (because focus on review). This a result from the survey, but in the theory we don't really touch upon process model. 
\subsubsection{RQ2: How do practitioners manage RED?}
The interviews presented in \Cref{sec:results:interview} show that the management of RED is composed of detecting, measuring, tracking, and remediating RED (P16). 
Detection is both the least expensive (P17) and most common (P18) type of interaction, likely because we lack specific tools for tracking (P19) and measuring RED (P21) in the industry.
The responsibility of addressing RED is perceived to lie with requirements engineers and high-level stakeholders (P22) and requires an explicit mandate (P23).

\subsubsection{Relevance to Research and Practice}

Technical debt is no longer limited to source code.
The metaphor has evolved to characterize circumstances where developers make compromises or take shortcuts throughout the development process, including when dealing with requirements. 
The practical manifestation and understanding of RED has an impact on both theory and practice. 

%Lack of product knowledge is caused by a number of circumstances, such as needs coming from several stakeholders (such as regulatory authorities) and rapid technological advancements that make it difficult to predict what characteristics a product should have after years of development (Liebelet et al. 2018). This inexperience may result in imprecise or ambiguous requirements (Liebelet et al. 2018). According to Liebel et al. (2018), decision rationales—which explain why and how a requirement was broken down to a more specific level—should be documented in order to prevent a mismatch between requirements on various abstraction levels. 

In particular for researchers, this work scopes a RE-specific theory to the vast TD literature and harmonizes the vocabulary used to address RED.
We see this empirically grounded theory as a fundamental basis to encourage and guide further RED research. 
The initial propositions serve as a set of falsifiable hypotheses, which invite follow-up, in-depth research on specific aspects of RED.
Furthermore, the theory can be expanded by adding further empirically grounded propositions or specializing existing propositions by contextualization; for example, whether the way RED items are tracked (P19) depends on specific Agile practices, such as backlog management. 
Finally, researchers are invited to address gaps confirmed by propositions, for example, by devising a method to reliably measure RED (E14).

Practitioners can use the resulting theory to compare their own experiences and practices against the provided propositions. 
Although in the current form the theory is not intended to be prescriptive, practitioners can reflect upon the corresponding explanations, contextualize them to their settings, and devise their own initiatives to manage RED.

\subsection{Relations to the Related Work and General Notion of TD}
\label{sec:discussion:mapping}

%The derived propositions focus on various RED related concepts such as causes, symptoms, value or roles. These propositions are based on the generalization of different occurrences of the concepts identified in literature, interviews and survey answers. 
We discuss how existing literature corroborates the propositions by highlighting categories and concept occurrences identified in this and other studies. 

Proposition 1 (P1) suggests that RED items can have time, people, artifact, and/or product-related causes. 
These categories were also identified in studies by Rios et al.~\cite{rios2020hearing} and Barbosa et al.~\cite{barbosaorganizing} that focused on documentation and requirements debt types, while the example occurrences of these categories are presented in \Cref{tab:conceptsliterature}.
Proposition 3 (P3) suggests that time-related causes are the most relevant causes of RED.
Time-related causes, conceptualized as \textit{deadline}, are also identified as the top-ranked cause for documentation and requirements type of debt in related studies~\cite{rios2020hearing, barbosaorganizing}.
Moreover, an impending \textit{deadline} is perceived as the most common cause of TD in general~\cite{ramac-jss}, indicating that the significance of deadline is also applicable in the RED domain. 

Similarly to proposition 1 (P1), proposition 10 (P10) suggests that RED items can have symptoms related to time, people, the product, and/or artifacts.
Available studies by Rios et al.~\cite{rios2020hearing} and Barbosa et al.~\cite{barbosaorganizing} corroborate this further.
This is also visible in the level of effect occurrences that form these categories. 
Proposition 11 (P11) suggests that the \textit{incompleteness of requirements} is the leading symptom of RED, while the following is the \textit{customer dissatisfaction}, as suggested in Proposition 12 (P12).
From \Cref{tab:conceptsliterature}, we observed that both previous studies~\cite{rios2020hearing, barbosaorganizing} support P12 by discussing \textit{low external quality}---interpreted as customer dissatisfaction---as one of the leading effects of documentation and requirement debt.
On the other hand, \textit{inadequate, non-existing or outdated documentation}, identified by Rios et al.~\cite{rios2020hearing}, directly supports the P11.
However, \textit{incompleteness of requirements}, or documentation of any kind, is not among the top 10 effects of TD in general, whereas the \textit{low external quality} occupies the fourth place~\cite{ramac-jss}. 
This shows a misalignment between the top-ranked effects of TD in general and the RED symptoms.
%Proposition 20 (P20)---the remediation strategy is strongly dependant on customer's perspective on the RED item---can also be supported with the study by Barbosa et al.~\cite{barbosaorganizing} which suggests that documentation debt repayment can sometimes be influenced by customers decision. 

\subsection{Threats to Validity}
\label{sec:discussion:threats}

To discuss the threats to validity, we adopt the guidelines proposed by Wohlin et al.~\cite{wohlin2012experimentation} and extended by Moll{\'e}ri et al.~\cite{molleri2020empirically}. 
% We report all  threats to validity applicable to our study.

\paragraph{External Validity} The online survey sample size limits the study result generalizability.
However, it still grants the proposal of an initial theory as a starting point for further development and evaluation as argued in \Cref{sec:results:theory}.
In addition, we further minimize this threat to validity by including demographic checks in the survey instrument to ensure the representativeness and eligibility of the participants (see~\Cref{tab:protocol:demographics}).

\paragraph{Conclusion Validity} We ensured the conclusion validity by carefully aligning the resulting RED theory with the established and more mature TD theory, when possible. 
The fourth to seventh authors reviewed the resulting propositions based on their involvement in TD research.
\reva{Confidence}{in our conclusion depends on the strength of the evidence found in the empirical studies---i.e., interviews and survey. We acknowledge that the results regarding development process models, such as Agile, are based, for the majority, on the answers to two survey questions.}
\paragraph{Internal Validity} Both the interview and the survey instrument were designed following established guidelines~\cite{molleri2020empirically}.
We piloted the instrument before deployment to ensure the understandability of the questions.
We piloted the survey among the first three authors and with one external senior reviewer to adjust the formulations and ensure unambiguous interpretations.

\paragraph{Interpretive Validity} Responses to open-ended survey questions were coded jointly by the first and second authors and validated by the third author to ensure a reliable inference of the participants' opinions.
However, a significant portion of responses was written in Brazilian Portuguese and translated using Google Translate. 
The fourth author, fluent in Brazilian Portuguese, confirmed the reliability of the translations by manually assessing a random sample.

% \paragraph{Other types of Validity} Participant- and researcher-targeted threats to validity were mitigated through diligent anonymization. 

\FloatBarrier
\section{Conclusion}
\label{sec:conclusion}
% summary
In this paper, we report on an initial theory of RED. 
Our theory includes propositions and explanations representing RED understanding and management in practice.
We developed the theory based on in-depth interviews with three experts in the area of RED and a global survey with 69 practitioners.  

% future work
The theory constitutes the first empirical foundation enabling future work on the debt metaphor in requirements engineering.
Contributions based on this theory will further the understanding of RED in practice by leveraging the debt metaphor to enable better decision-making in RE.
Further research needs to synthesize the insights gained on RED, and the problems and challenges practitioners face in RE, as revealed in the Naming the Pain in Requirements Engineering initiative (see~\url{www.napire.org}).
Such endeavor will allow us to explore further root causes for RED as well as better understand its impact.
Additional empirical work must verify or refute the existing propositions and contextualize and extend them.  
\revb{Such}{endeavor will allow us to further explore root causes for and impacts of RED and will provide a more robust basis for further reasoning about phenomena surrounding RED.
These include the notion of intentionality in connection to \textit{laziness} or \textit{upfront lack of commitment to RE}, which has not been discussed in the scope of this study, or the impact of the \textit{software process model} (like Agile), which deserves more attention.}

\section*{Acknowledgements}
This work was supported by the KKS foundation through the S.E.R.T. Research Profile project at Blekinge Institute of Technology. \reva{We}{additionally thank the reviewers for their valuable feedback} \revb{that}{helped strengthen the manuscript.}

\bibliography{material/references}

\newpage
\appendix

\section{Survey Protocol}
\label{sec:apx:survey}

Questions are marked as single choice (SC) or multiple choice (MC). Single-choice questions answered on the five-point Likert scale are marked as (SCL). Open questions are marked as (O).

Several questions asking respondents to name stakeholders use the following list: Requirements Engineer/Business analyst, product owner, system/software architect, developer, tester, quality engineer/manager, trainer/educator, researcher in industry, manager, other (please specify). This list is hereby referred to as (stakeholders).

\subsection{Understanding requirements engineering debt}

\begin{table}[t]
    \centering
    \begin{tabular}{c|p{9.4cm}|c}
        ID & Question - Options & Type \\ \toprule
        Q1 & In which country are you working? - (list of countries) & SC \\
        Q2 & Which primary job functions did you have in the last years? - (stakeholders) & SC \\
        Q3 & Where do you identify your profession on a scale between research and practice? & SCL \\
        Q4 & How many years of general experience do you have from working with requirements in any form? & O \\
        Q5 & How many people are involved in the latest project you have been working on? & O \\
        Q6 & Which organisational role does your project team have in your company? - Main contractor (main responsible for the development), sub-contractor (responsible for part of a larger development project), in-house development, other (please specify) & SC \\
        Q7 & Please select the main industrial sector of your project and the application domain of the systems you build. - agriculture, automotive, finance, healthcare, security, manufacturing, energy, logistics, railway, avionics, insurance, education, public sector, enterprise resource planning, human resources, e-Government, telecommunication, games engineering, public transportation, e-Commerce, other (please specify) & SC \\
    \end{tabular}
    \caption{Survey Protocol: Demographic Questions}
    \label{tab:protocol:demographics}
\end{table}

\begin{table}[t]
    \centering
    \begin{tabular}{c|p{9.4cm}|c}
        ID & Question - Options & Type \\ \toprule
        Q8 & In your experience, which of the following options qualify as causes of requirements engineering debt? - taking shortcuts while fixing requirements, business decisions, lack of (domain) knowledge, lack of diligence, tradeoff between cost and benefit, tradeoff between different requirements, requirements documentation in artefacts, information scattered across artefacts, information scattered across stakeholders, lack of requirements elicitation process, inconsistencies between different visions for the same product, lack of tooling support for RE, changes in the market of the product, requirements are only specified at a high level, technical complexity of the project, time pressure to deliver a feature, requirements engineers lack awareness of requirements debt, lack of communication between stakeholders, lack of prioritization, changes in requirements documentation, lack of requirements formalization, specifying cross-cutting features, requirements not documented, lack of requirements traceability, time pressure to finalize the (systematic) requirements specification, other (please specify) & MC \\
        Q9 & To what extent do you agree with the statement: ``The overall software process model has an influence on the likelihood of introducing requirements engineering debt.'' & SCL \\
        Q10 & Please motivate your answer to the question above. & O \\
    \end{tabular}
    \caption{Survey Protocol: Causes}
    \label{tab:protocol:causes}
\end{table}

\begin{table}[t]
    \centering
    \begin{tabular}{c|p{9.4cm}|c}
        ID & Question - Options & Type \\ \toprule
        Q11 & What of the following qualifies as the value of introducing requirements engineering debt? - faster time to market, advantage over competitors, focus on feature with better return on investment, legal obligations, decisions based on authorative opinions, other (please specify) & MC \\
        Q12 & In your projects, do you calculate the value of requirements debt? & SC \\
        Q13 & Who estimates the value of requirements engineering debt? - entire company, (stakeholders) & MC \\
        Q14 & In your opinion, which factors influence these values? & O \\
        Q15 & How is this value calculated? & O \\
    \end{tabular}
    \caption{Survey Protocol: value}
    \label{tab:protocol:value}
\end{table}

\begin{table}[t]
    \centering
    \begin{tabular}{c|p{9.4cm}|c}
        ID & Question - Options & Type \\ \toprule
        Q16 & In your experience, which of the following options qualify as a symptom of requirements engineering debt? - not implementing requirements, disregarding automation processes, architectural cost, lack of usage of the product, additional operational cost of the product, decreased reputation of the company/product, sow release of a product, incompleteness of requirements, not involving all relevant stakeholders, others (please specify) & MC \\
        Q17 & To what extent do you agree with the statement: ``Requirements engineering debt can cause a product to go bankrupt.'' & SCL \\
        Q18 & Rate your agreement to the statement, that the following symptoms of RED can lead to bankruptcy. (rating all options from Q16) & SCL \\
        Q19 & Do you degree with the statement: "Requirements engineering debt can cause a single requirements specification to bankrupt?" & SCL
    \end{tabular}
    \caption{Survey Protocol: Symptoms}
    \label{tab:protocol:symptoms}
\end{table}

\begin{table}[t]
    \centering
    \begin{tabular}{c|p{9.4cm}|c}
        ID & Question - Options & Type \\ \toprule
        Q20 & To what extent do you agree with the statement: ``Requirements engineering debt can be introduced intentionally''? & SCL \\
        Q21 & To what extent do you agree with the statement: ``Requirements engineering debt can be introduced unintentionally''? & SCL \\
        Q22 & Can you elaborate on your perception of intentionality for requirements engineering debt? & O \\
    \end{tabular}
    \caption{Survey Protocol: Intentionality}
    \label{tab:protocol:intentionality}
\end{table}

\begin{table}[t]
    \centering
    \begin{tabular}{c|p{9.4cm}|c}
        ID & Question - Options & Type \\ \toprule
        Q23 & Which of the following context factors increase the likelihood of propagation of requirements engineering debt? - maturity of the product under development (PUD), quality of the PUD, size of the PUD, type of the PUD, customization of the PUD, development process, type of artefacts involved in the development, software development tools and techniques, Other practices and techniques not related to the development process, roles implemented in the company, experience of the people involved, model of the overall organization, distribution of the organization, company strategy in the market, constraints by the market & SCL \\
        Q24 & In your opinion, what are the most prominent effects of requirements engineering debt on software development activities and artefacts? & O \\
    \end{tabular}
    \caption{Survey Protocol: Propagation}
    \label{tab:protocol:propagation}
\end{table}

\subsection{Managing requirements engineering debt}

\begin{table}[t]
    \centering
    \begin{tabular}{c|p{9.4cm}|c}
        ID & Question - Options & Type \\ \toprule
        Q25 & To what extent do you agree with the statement: ``Detecting requirements engineering debt is expensive.'' & SCL \\
        Q26 & Do you explicitly detect requirements engineering debt in your company/project? & SC \\
        Q27 & When do you tend to identify requirements engineering debt? - during the requirements validation, explicitly before eliciting requirements for a new feature/release, during the overall risk assessment of the project, other (please specify) & SC \\
        Q28 & In your opinion, what techniques can be utilized for detecting requirements engineering debt? & O \\
        Q29 & Who is responsible for detecting requirements engineering debt? - entire company, (stakeholders) & SC \\
    \end{tabular}
    \caption{Survey Protocol: Detecting}
    \label{tab:protocol:causes}
\end{table}

\begin{table}[t]
    \centering
    \begin{tabular}{c|p{9.4cm}|c}
        ID & Question - Options & Type \\ \toprule
        Q30 & To what extent do you agree with the statement: ``Measuring requirements engineering debt is expensive.'' & SCL \\
        Q31 & Do you explicitly measure requirements engineering debt in your company/project? & SC \\
        Q32 & In your experience, what techniques are eligible for measuring requirements engineering debt? & O \\
        Q33 & Who is responsible for measuring requirements engineering debt? - entire company, (stakeholders) & SC \\
    \end{tabular}
    \caption{Survey Protocol: Measuring}
    \label{tab:protocol:measuring}
\end{table}

\begin{table}[t]
    \centering
    \begin{tabular}{c|p{9.4cm}|c}
        ID & Question - Options & Type \\ \toprule
        Q34 & To what extent do you agree with the statement: ``Tracking requirements engineering debt is expensive.'' & SCL \\
        Q35 & Do you explicitly track requirements engineering debt in your company/project? & SC \\
        Q36 & In your experience, what techniques are eligible for tracking requirements engineering debt? & O \\
        Q37 & Who is responsible for tracking requirements engineering debt? - entire company, (stakeholders) & SC \\
    \end{tabular}
    \caption{Survey Protocol: Tracking}
    \label{tab:protocol:tracking}
\end{table}

\begin{table}[t]
    \centering
    \begin{tabular}{c|p{9.4cm}|c}
        ID & Question - Options & Type \\ \toprule
        Q38 & To what extent do you agree with the statement: ``Remediating requirements engineering debt is expensive.'' & SCL \\
        Q39 & Do you explicitly remediate requirements engineering debt in your company/project? & SC \\
        Q40 & In your experience, what techniques can be utilized for remediating requirements engineering debt? & O \\
        Q41 & To what extent do you agree with the statement: ``Introducing new debt items precludes clearing old debt items''? & SCL \\
        Q42 & Who is responsible for remediating requirements engineering debt? - entire company, (stakeholders) & SC \\
        Q43 & To what extent do you agree with the statement: ``Remediating requirements engineering debt requires a mandate''? & SCL \\
    \end{tabular}
    \caption{Survey Protocol: Remediating}
    \label{tab:protocol:remediating}
\end{table}

\FloatBarrier
\section{Answer categories}
\label{sec:apx:categories}
For several survey results reported in Section~\ref{sec:results}, the answers were grouped by categories. The following tables describe all groups and contained answers for causes (Table~\ref{tab:apx:groups:cause}), values (Table~\ref{tab:apx:groups:value}), symptoms (Table~\ref{tab:apx:groups:symptoms}), and context factors~\cite{petersen2009context} (Table~\ref{tab:apx:groups:context})

\begin{table}[ht]
\footnotesize
    \centering
    \begin{tabular}{>{\centering\bfseries}m{1cm}  >{\centering\arraybackslash}m{9.7cm}}
    \toprule
        \textbf{Category} & \textbf{Answers} \\ \midrule
        Time & 
         \renewcommand\labelitemi{\tiny$\bullet$}
        \begin{itemize}
         \setlength\itemsep{-0.25cm}
            \item Taking shortcuts while fixing requirements
            \item Time pressure to deliver a feature
            \item Time pressure to finalize the (systematic) requirements specification 
        \end{itemize} \\ [-0.4cm]
        \midrule
        Product & 
         \renewcommand\labelitemi{\tiny$\bullet$}
        \begin{itemize}
         \setlength\itemsep{-0.25cm}
            \item Business decisions
            \item Trade-off between cost and benefit
            \item Changes in the market of the product
            \item Technical complexity of the product
            \item Specifying cross-cutting features  
        \end{itemize}
        \\[-0.4cm]
        \midrule
        People & 
         \renewcommand\labelitemi{\tiny$\bullet$}
        \begin{itemize}
         \setlength\itemsep{-0.25cm}
            \item Lack of (domain) knowledge
            \item Information scattered across stakeholders
            \item Inconsistencies between different visions for the same product 
            \item Requirements engineers lack awareness of requirements debt
            \item Lack of communication between stakeholders  
        \end{itemize}
        \\ [-0.4cm] 
        \midrule
        Artifact & 
         \renewcommand\labelitemi{\tiny$\bullet$}
        \begin{itemize}
        \setlength\itemsep{-0.25cm}    
            \item Trade-off between different requirements
            \item Requirements documentation in artefacts
            \item Information scattered across artefacts
            \item Lack of requirements elicitation process 
            \item Lack of tooling support for RE 
            \item Requirements are only specified at a high level
            \item Lack of prioritization
            \item Changes in requirements documentation
            \item Lack of requirements formalization
            \item Requirements not documented
            \item Lack of requirements traceability 
        \end{itemize} \\ [-0.4cm]
        \bottomrule
    \end{tabular}
    \caption{Groups of answers for causes}
    \label{tab:apx:groups:cause}
\end{table}

\begin{table}[ht]
    \centering
    \begin{tabular}{c|p{9.5cm}} \toprule
        \textbf{Category} & \textbf{Answers} \\ \midrule
        Time & Faster time to market \\
        Product & Advantage over competitors, focus on feature with better return on investment, legal obligations \\
        People & Decisions based on authoritative opinions \\
        \bottomrule
    \end{tabular}
    \caption{Groups of answers for values}
    \label{tab:apx:groups:value}
\end{table}

\begin{table}[ht]
    \centering
    \begin{tabular}{c|p{9.5cm}} \toprule
        \textbf{Category} & \textbf{Answers} \\ \midrule
        Product & Not implementing requirements, architectural cost, lack of usage of the product, additional operational cost of the product, decreased reputation of the company/product, slow release of a product \\
        People & Disregarding automation processes, not involving all relevant stakeholders \\
        Artifact & Incompleteness of requirements \\ 
        \bottomrule
    \end{tabular}
    \caption{Groups of answers for symptoms}
    \label{tab:apx:groups:symptoms}
\end{table}

\begin{table}[ht]
    \centering
    \begin{tabular}{c|p{9.5cm}} \toprule
        \textbf{Category} & \textbf{Answers} \\ \midrule
        Product & Maturity of the product under development, Quality of the product under development, Size of the product under development, System type of the product under development, Customization of the product under development \\
        Process & Activities of the development process, Artefacts of the development process \\
        Tools & Usage of tools for automated software development, Other practices and techniques not related to the development process \\
        People & Roles implemented in the company, Experience of people involved, Model of overall organization, Distribution of the organization, Strategy of addressing the market \\
        Market & Constraints by the market \\
        \bottomrule
    \end{tabular}
    \caption{Groups of answers for context factors}
    \label{tab:apx:groups:context}
\end{table}

\newpage
\section*{CRediT}
We report the contribution of the authors to this manuscript using the Contributor Roles Taxonomy (CRediT) by Larivi{\`e}re et al.~\cite{lariviere2021CRediT}:
\begin{itemize}
    \item \textbf{Julian Frattini}: Conceptualization (equal), Data curation (lead), Investigation (equal), Methodology (equal), Visualization (lead), Project administration (lead),  Writing - Original Draft Preparation (lead), Writing - Review \& Editing (supporting)
    \item \textbf{Davide Fucci}: Conceptualization (equal), Data curation (supporting), Investigation (equal), Resources (equal), Methodology (equal), Project administration (supporting), Supervision (lead), Writing – Original Draft Preparation (supporting), Writing – review \& editing (equal).
    \item \textbf{Daniel Mendez}: Conceptualization (supporting), Funding acquisition (lead), Investigation (equal), Methodology (supporting), Resources (lead), Supervision (supporting), Writing – Original Draft Preparation (supporting), Writing – review \& editing (supporting). 
    \item \textbf{Rodrigo Spinola}: Resources (equal), Validation (supporting),  Writing - Review \& Editing (supporting)
    \item \textbf{Vladimir Mandi\'c}: Resources (equal),  Writing - Review \& Editing (supporting)
    \item \textbf{Neboj\v sa Tau\v san}: Resources (equal), Writing - Original Draft Preparation (supporting),  Writing - Review \& Editing (supporting)
    \item \textbf{Muhammad Ovais Ahmad}: Resources (equal),  Writing - Original Draft Preparation (supporting), Writing - Review \& Editing (supporting)
    \item \textbf{Javier Gonzalez-Huerta}: Conceptualization (supporting),  Writing - Review \& Editing (supporting)

\end{itemize}
\end{document}